\documentclass[acmsmall,screen]{acmart}

\usepackage{graphicx}
\usepackage{tikz}
\usepackage{graphicx}
\usetikzlibrary{arrows.meta,positioning,fit,backgrounds,calc,shapes.geometric,shapes.arrows,shapes.multipart}
\usepackage[edges]{forest}
\usepackage{adjustbox}
\usepackage{xcolor}
\usepackage{amsmath}
\usepackage{ragged2e}
\usepackage{multirow}
\usepackage{multicol}
\usepackage{array}
\usepackage{enumitem}
\usepackage{makecell}

\AtBeginDocument{%
  }

\begin{document}

\title{Large Language Models in Misinformation Ecosystems: Misuse, Defense, and Vulnerability}

\author{Lingwei Wei}
\orcid{0000-0002-7058-2662}
\affiliation{%
  \institution{Institute of Information Engineering, Chinese Academy of Sciences}
  \city{Beijing}
  \state{}
  \country{China}
}
\affiliation{%
 \institution{University of Illinois Chicago}
 \city{Chicago}
 \state{Illinois}
 \country{USA}
 } 
\email{weilingwei@iie.ac.cn}

\author{Dou Hu}
\orcid{0000-0001-7790-8568}
\affiliation{%
  \institution{State Key Laboratory of Media Convergence and Communication, Communication University of China}
  \city{Beijing}
    \state{}
\country{China}}
  \email{hudou@cuc.edu.cn}

\author{Wei Zhou}
\orcid{0000-0003-3622-3970}
\affiliation{%
  \institution{Institute of Information Engineering, Chinese Academy of Sciences}
  \city{Beijing}
    \state{}
\country{China}
  }
\email{zhouwei@iie.ac.cn}

\author{Songlin Hu}
\orcid{0000-0002-7170-3809}
\affiliation{%
 \institution{Institute of Information Engineering, Chinese Academy of Sciences}
  \city{Beijing}
   \state{}
 \country{China}
  }  
\affiliation{ 
  \institution{University of Chinese Academy of Sciences}
  \city{Beijing}
   \state{}
 \country{China}
}
\email{husonglin@iie.ac.cn}

\author{Philip S. Yu}
\orcid{0000-0002-3491-5968}
\affiliation{%
 \institution{University of Illinois Chicago}
 \city{Chicago}
 \state{Illinois}
 \country{USA}}
\email{psuyu@uic.edu}
 
\setcopyright{none}
\settopmatter{printacmref=false}
\renewcommand\footnotetextcopyrightpermission[1]{}

\begin{abstract}
Large language models (LLMs) have transformed misinformation from a primarily content-centric problem into a broader ecosystem-level security challenge. When misused, LLMs create risks beyond false content generation, enabling attacks on the social contexts, evidence sources, retrieval corpora, and verification workflows that misinformation defense depends on. In this paper, we introduce a role-layer framework to unify these risks and defenses. The role dimension characterizes LLMs as attackers, defenders, and vulnerable components of verification systems, while the layer dimension covers content, social contexts, evidence environments, and verification workflows. Guided by this framework, we organize LLM-enabled attacks, investigate LLM-based detection and verification methods, analyze vulnerabilities in LLM-centric detection paradigms, and discuss existing countermeasures against LLM-enabled attacks. Building on this synthesis, we identify three key open challenges: moving from static detection accuracy to budgeted ecosystem-level risk evaluation, hardening LLM-centered verification pipelines against adversarial manipulation, and deploying auditable human-in-the-loop verification systems for trustworthy real-world misinformation defense.
\end{abstract}


\maketitle

\section{Introduction}
Misinformation, broadly defined as false or misleading information regardless of intent~\cite{zhou2020survey,aimeur2023fake,chen2023spread,hartwig2024landscape},  has long posed a major threat to the information ecosystem.  
The emergence of large language models (LLMs) has substantially reshaped this threat landscape.
LLMs lower the cost of producing fluent, coherent, and seemingly authoritative misinformation \cite{han2025exploring}. They can generate fabricated news articles, misleading summaries, persuasive arguments, false explanations, and localized narratives within seconds. 
More importantly, LLMs amplify misinformation risks beyond the generation of synthetic false content. 
Recent studies have also shown that LLMs can be misused to rewrite existing claims to evade detectors \cite{park2025adversarial}, personalize narratives for different audiences \cite{nasiri2025evolution}, translate and localize false information across languages \cite{kaneko2026jailnewsbench}, simulate personas \cite{farr2025simulating}, generate comments \cite{luo2024message} and social interactions \cite{lu2025understanding}, and assist coordinated manipulation campaigns \cite{chen2026networked}.
As a result, LLMs do not merely increase the volume of misinformation; they also broaden the threat surface by enabling misinformation to be created, amplified, contextualized, and concealed in previous impossible ways.

At the same time, LLMs are increasingly used as defensive components in misinformation detection and mitigation pipelines. Recent work explores LLM-based detectors~\cite{tian2025llm}, retrieval-augmented generation (RAG) systems~\cite{zhao2024optimizing}, multimodal fact-checkers~\cite{yang2026rama}, and agentic verification pipelines~\cite{li2024large_ecai,muneer2026mosaiv} for misinformation detection and mitigation. Compared with conventional detection systems, these LLM-enabled approaches support open-ended reasoning, evidence-grounded verification, natural-language explanation, and tool-assisted fact-checking. This line of work highlights the defense role of LLMs in the misinformation ecosystem, i.e., LLMs can assist defenders by expanding the capabilities of detection and verification systems.

Although LLMs create new opportunities for scalable defense, they also introduce new vulnerabilities, making these pipelines themselves targets of adversarial manipulation. For example, a RAG-based fact-checking system may produce misleading conclusions if its retrieval corpus is poisoned or adversarial documents are ranked as supporting evidence~\cite{zou2025poisonedrag,ha2025mm}. An LLM-based verifier may be misled by prompt injection~\cite{leite2026llm}. An agentic detection system that browses the web, invokes tools, or coordinates multiple agents may further propagate errors across multiple reasoning and action steps~\cite{tian2024web,cui2025toward}. 
Thus, LLMs occupy a third role, i.e., they become vulnerable components in misinformation detection and mitigation systems.

\begin{table}[t]
\centering
\small
\caption{Comparison of existing LLM-related papers on misinformation detection with our work.}
\label{tab:comparison}
\resizebox{\linewidth}{!}{
\begin{tabular}{l|c| ccc ccc }
\toprule
\multicolumn{1}{c|}{\multirow{2}{*}{Paper}}  &   \multicolumn{1}{c|}{\multirow{2}{*}{Year}}    
& \multicolumn{3}{c}{LLM as Attacker}  & \multicolumn{2}{c}{\multirow{1}{*}{LLM as Defender}}  & \multicolumn{1}{c}{\multirow{1}{*}{LLM as Victim}} 
\\ 
&  & Content & Social & Evidence &  Detection & Mitigation &   
\\
\midrule
\citet{chen2024combating} & 2024 & $\checkmark$ & $\times$ &  $\times$ & $\checkmark$ &  $\times$ & $\times$  \\ 
\citet{papageorgiou2024survey} & 2024 &  $\checkmark$ & $\times$ &  $\times$ & $\checkmark$ &  $\times$ & $\times$  \\ 
\citet{lucas2024longtail} & 2024 & $\checkmark$ & $\times$ &  $\times$ & $\times$ &  $\times$ &  $\checkmark$   \\ 
\citet{yi2025challenges} & 2025 & $\times$ &  $\times$  &  $\times$ & $\checkmark$ & $\times$  & $\times$  \\ 
\citet{liu2025survey} & 2025 & $\times$ &  $\times$  &  $\times$ & $\checkmark$ & $\checkmark$  & $\times$  \\ 
\citet{xie2025survey} & 2025 & $\times$ & $\times$ & $\times$ & $\checkmark$ & $\checkmark$ & $\times$ \\ 
\citet{liu2025adversarial} & 2025 & $\checkmark$ & $\times$  & $\checkmark$ & $\checkmark$ & $\checkmark$ & $\times$  \\ 
\citet{yang2025rethink} & 2025 & $\times$ & $\times$ &  $\times$ & $\checkmark$ &  $\checkmark$ &  $\times$   \\ 
\citet{park2026generative} & 2026 & $\checkmark$ & $\times$ &  $\times$ & $\checkmark$ & $\checkmark$  & $\times$  \\
\citet{jainadversarial2026} & 2026 & $\times$ & $\times$ &  $\times$ & $\times$ & $\times$  & $\checkmark$  \\  
\textbf{Ours} &  2026 & $\checkmark$  & $\checkmark$  & $\checkmark$  & $\checkmark$  & $\checkmark$ & $\checkmark$  \\
\bottomrule
\end{tabular}
}
\end{table}

\begin{figure}[t]
    \centering
    \includegraphics[width=0.6\linewidth]{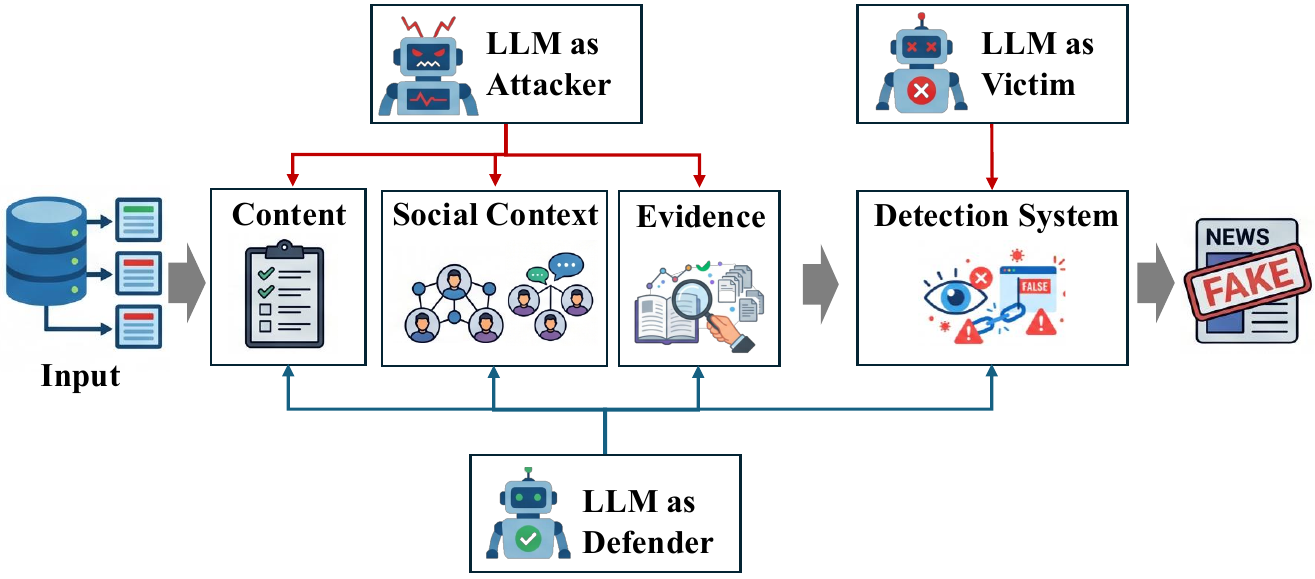}
    \caption{Overview of LLM roles in the misinformation ecosystem.}
    \label{fig:overview}
\end{figure}

These developments suggest that LLMs play multiple roles in the misinformation ecosystem: they can empower attackers, assist defenders, and become vulnerable components within detection and mitigation systems.
However, previous related works often focus on one particular aspect of the problem, such as LLM-generated fake content, LLM-based detection, internal LLM defense, propagation modeling, or fact-checking robustness. 
Table~\ref{tab:comparison} summarizes these works according to the roles played by LLMs in the misinformation ecosystem, including LLMs as attackers, defenders, and victims.
Some studies examine misinformation in the age of LLMs and discuss the opportunities and challenges brought by LLMs for misinformation detection and mitigation~\citep{chen2024combating,papageorgiou2024survey,park2026generative}. Others focus on LLM-powered fake news detection and proactive defense against misinformation in LLMs~\citep{yi2025challenges,liu2025survey}. Recent robustness-oriented works further investigate adversarial attacks against automated fact-checking systems~\citep{liu2025adversarial} or vulnerabilities of multimodal large language models~\citep{jainadversarial2026}. 
Therefore, a unified view is still missing on how LLMs simultaneously empower attackers, assist defenders, and become vulnerable components across the misinformation ecosystem.

\begin{table*}[t]
\centering
\small
\caption{Role-layer framework for LLMs in the misinformation ecosystem. }
\label{tab:two_axis_framework}
\resizebox{\linewidth}{!}{
\begin{tabular}{p{1.6cm} p{3.2cm} p{3.8cm} p{5.0cm}}
\toprule
\textbf{LLM Role} & \textbf{Core Question} & \textbf{Layered Focus} & \textbf{Representative Focus} \\
\midrule
\textbf{Attacker} (Sec.~\ref{sec:attacker})
& How do LLMs empower misinformation manipulation?
& Content, social context, evidence environment
& Claim generation and rewriting, persona and comment simulation, evidence fabrication, corpus poisoning, retrieval manipulation
\\
\midrule
\textbf{Victim} (Sec.~\ref{sec:victims})
& How do LLM-centric verification pipelines fail under adversarial manipulation?
& Verification workflow, with dependencies on content, social context, and evidence environments
& LLM-as-judge failures, RAG vulnerabilities, agentic workflow compromise 
\\
\midrule
\textbf{Defender} (Sec.~\ref{sec:defender})
& How do LLMs support detection, verification, and mitigation?
& Content, social context, evidence, system workflow
& LLM-based detection, social-context analysis, evidence-grounded verification, secure RAG, agent auditing \\
\bottomrule
\end{tabular}
}
\end{table*}

To address this gap, we introduce a role-layer framework to organize LLM-related misinformation research, as shown in Table~\ref{tab:two_axis_framework}.  
The role dimension distinguishes whether LLMs are used as attackers, defenders, or vulnerable components of detection systems. In the attacker role, LLMs are misused to generate, amplify, or strategically optimize misinformation. In the defender role, LLMs are utilized to support detection, mitigation, and explanation. In the victim role, the vulnerable object is the LLM-centered detection pipeline, including LLM-as-judge detectors, and agentic detection workflows.
The layer dimension characterizes where manipulation, vulnerability, or defense emerges. We distinguish four layers: content, social contexts, evidence environments, and verification workflows. The content layer focuses on what is stated, such as claim and news articles. The social layer concerns the propagation environment in which content is shared and amplified such as comments, user interactive behaviors. The evidence layer captures the external information sources to support verification, such as knowledge bases. The verification workflows refer to the system-level processes, which are not only defensive mechanisms but also potential targets of adversarial manipulation.
This role-layer framework integrates LLM-enabled attacks, system vulnerabilities, and defensive methods within a unified structure.

\begin{figure}[t]
    \centering
    \includegraphics[width=0.84\linewidth]{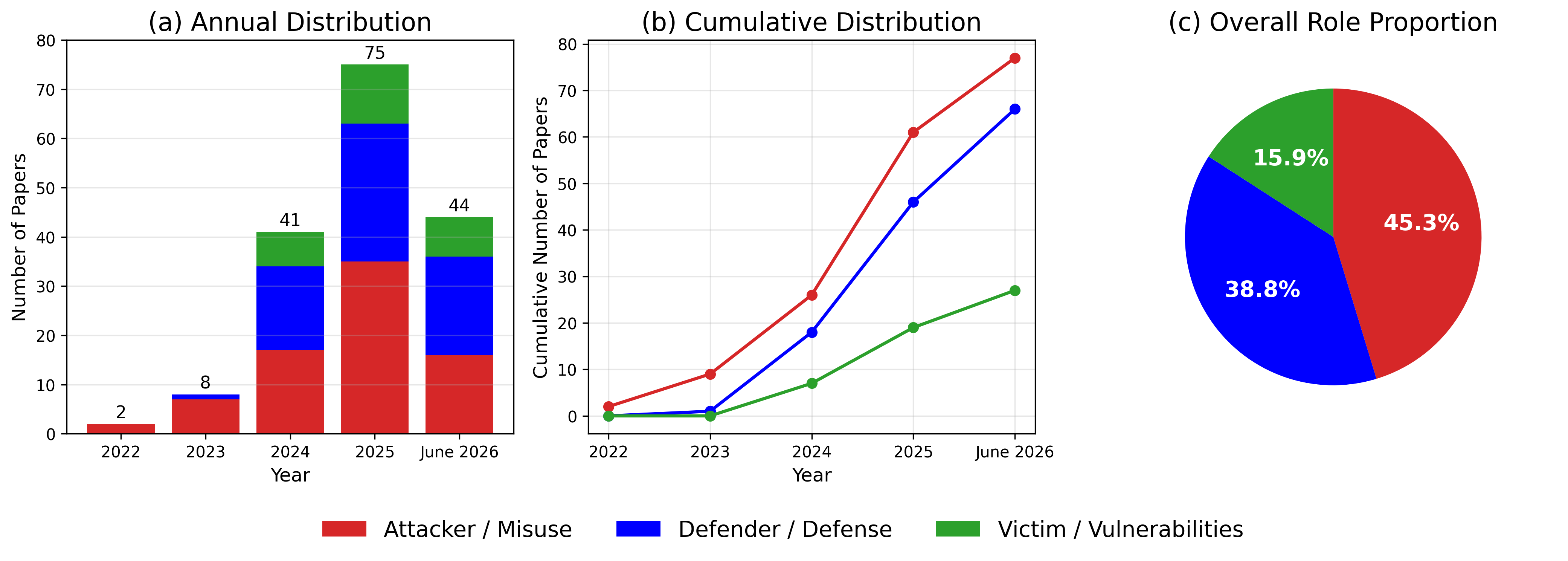}
   \caption{Temporal distribution of references across the three LLM roles from 2022 to June 2026 in this work.}
\label{fig:temporal_role_distribution}
\end{figure}

Following this framework, we organize around the tri-role perspective.
Fig.~\ref{fig:temporal_role_distribution} summarizes the temporal distribution of references across the three roles from 2022 to June 2026. The results show that research on LLM misuse has grown rapidly after 2023, reflecting increasing attention to LLM-enabled attacks. LLM-based defense studies also increase substantially, especially after 2024, whereas research treating LLMs as vulnerable components remains comparatively limited but has shown a clear recent rise. 
Section~\ref{sec:attacker} summarizes how LLMs empower attackers across content, social, and evidence layers. Section~\ref{sec:victims} then discusses how detection and mitigation systems built around LLMs can be compromised by adversarial claims, poisoned evidence, prompt injection, and agentic tool-use manipulation. Section~\ref{sec:defender} turns to the defensive use of LLMs, focusing on their roles in content understanding and generation, judge-based and agentic misinformation detection, and retrieval-augmented verification. 
Section~\ref{sec:countermeasures} further analyzes robustness-oriented countermeasures against LLM-enabled misinformation risks.

This paper makes the following main contributions:
\begin{enumerate}
\item We introduce a role-layer framework to organize LLM-related misinformation research along two complementary dimensions: the role of LLMs in the misinformation ecosystem including attacker, defender, and victim perspectives, and the layer at which manipulation, vulnerability, or defense arises including content, social contexts, evidence, and verification workflows.
\item We organize LLM-enabled threats beyond synthetic content generation, showing how LLM misuse expands misinformation risks across content, social, and evidence surfaces.
\item We systematically analyze LLM-enabled defenses and robustness-oriented countermeasures, thereby revealing how defensive uses of LLMs correspond to different attack surfaces.
\item We further discuss the underexplored vulnerability of LLM-centric misinformation detection systems, highlighting how LLM-as-judge detectors and agentic fact-checking workflows can be attacked in adversarial information environments.
\item We identify three key open challenges: transitioning from static detection accuracy to budget-aware ecosystem-level risk evaluation, hardening LLM-based detection pipelines against adversarial manipulation, and deploying auditable human-in-the-loop verification systems for trustworthy real-world misinformation defense.
\end{enumerate}

\section{LLMs as Attackers: Empowered Misinformation Attacks} \label{sec:attacker}

\begin{figure}[t]
    \centering
    \includegraphics[width=0.7\linewidth]{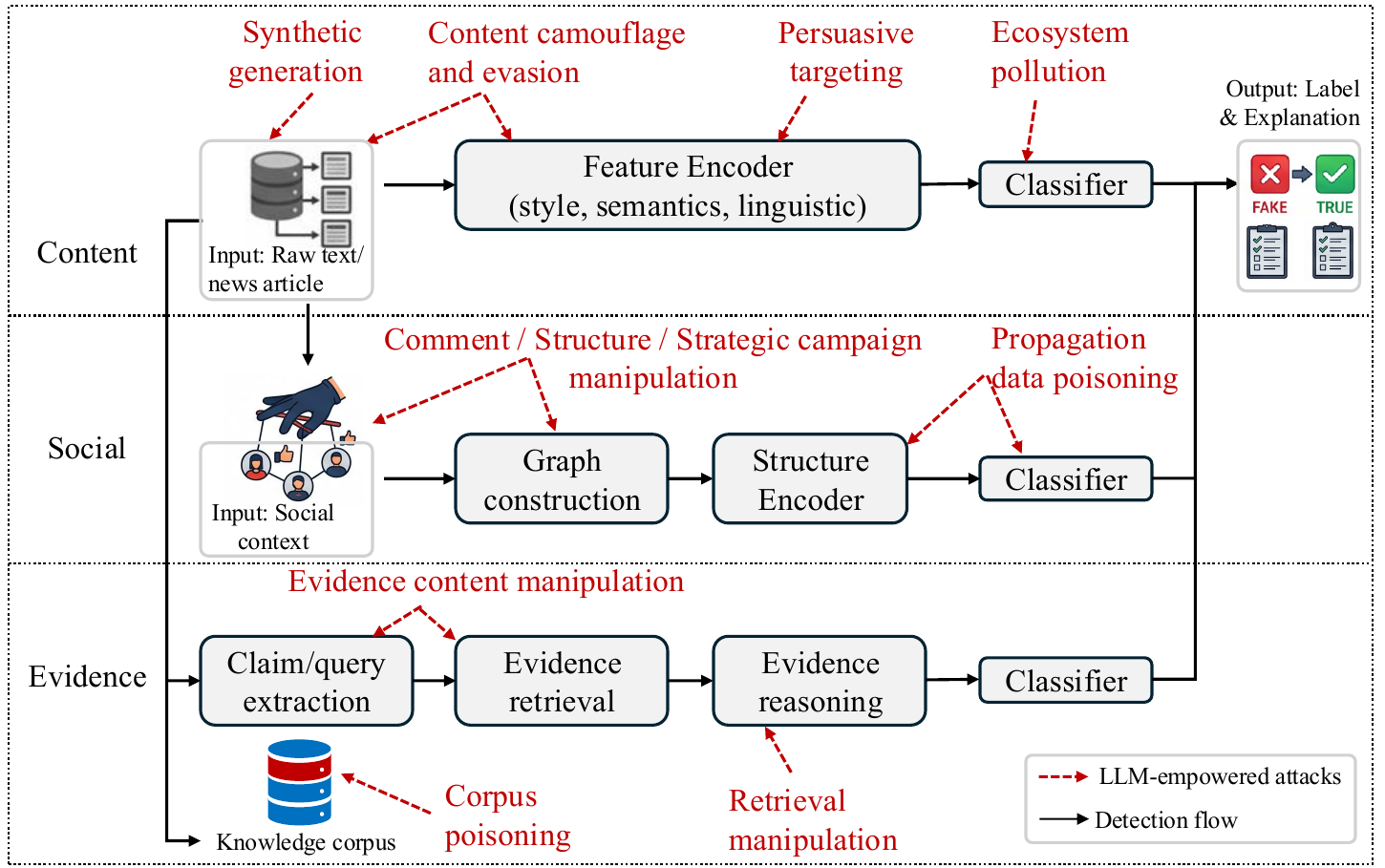}
    \caption{Overview of LLM-empowered attack for misinformation detection.}
    \label{fig:attack_overview}
\end{figure}

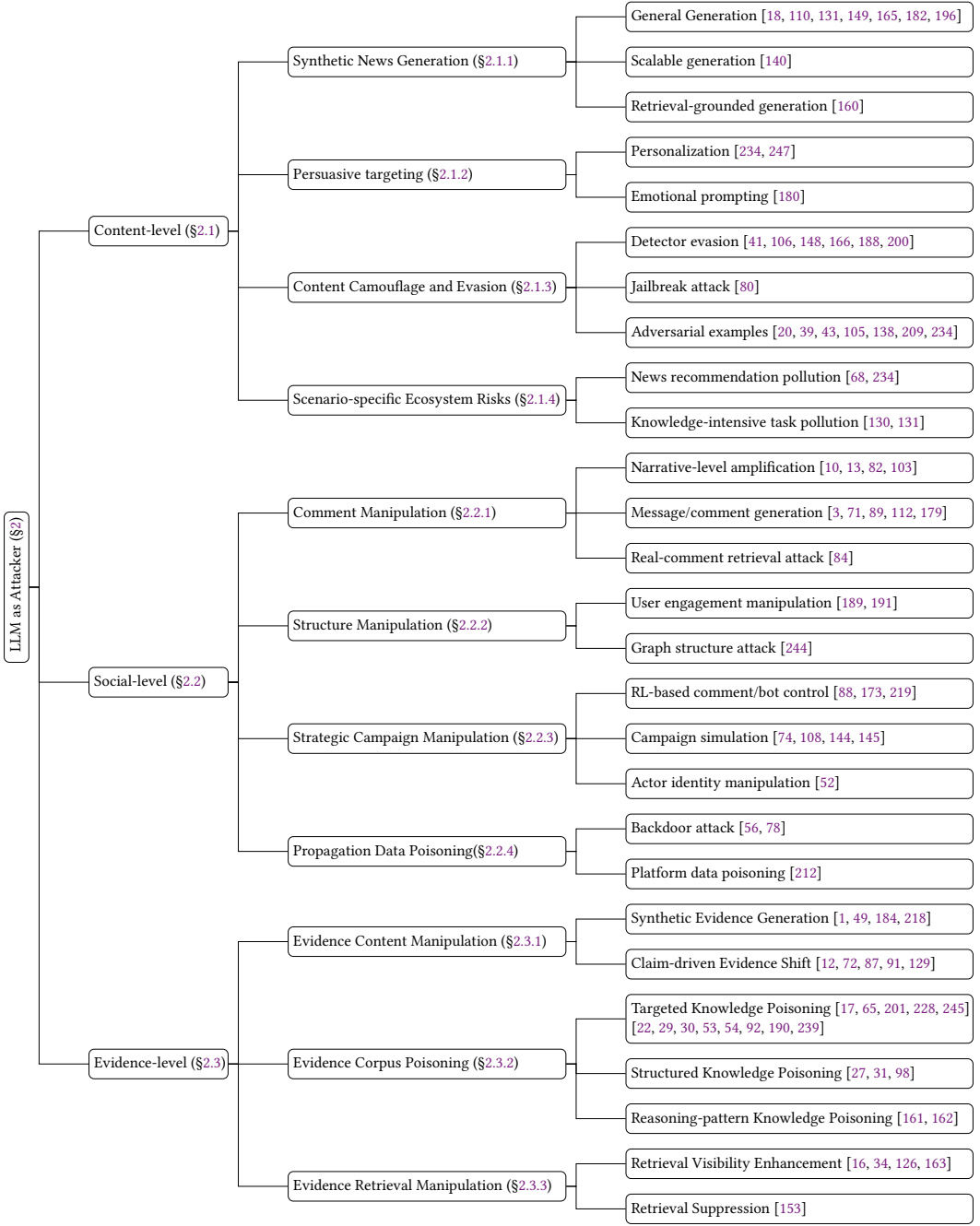
\begin{figure*}[t]
\centering
\begin{adjustbox}{max width=\textwidth, max height=0.88\textheight, center}
\begin{forest}
for tree={
    grow'=east,
    draw,
    rounded corners=3pt,
    align=left,
    font=\small,
    inner sep=3pt,
    edge={draw, -},
    parent anchor=east,
    child anchor=west,
    l sep=12mm,
    s sep=3mm,
    forked edges,
    fork sep=2mm,
    execute at begin node=\RaggedRight\sloppy,   
}
[\rotatebox{90}{LLM as Attacker (\S \ref{sec:attacker})}, 
    [Content-level (\S \ref{sec:attack_content}),
        text width=26mm
        [Synthetic News Generation (\S \ref{sec:attack_content_1}), 
            text width=54mm
            [General Generation 
            \cite{pan2023risk,lucas2023fighting,su2023fake,vykopal2024disinformation,sallami2024deception,chen2024can,wang2025have},
                text width=68mm,
                align=left
            ]
            [Scalable generation \cite{puccetti2024ai} \\ ,
                text width=68mm,
                align=left
            ]
            [Retrieval-grounded generation \cite{singh2024adversarial},
                text width=68mm,
                align=left
            ]
        ]
        [Persuasive targeting (\S \ref{sec:attack_content_2}), 
            text width=54mm   
            [Personalization \cite{zugecova2025evaluation,zhao2025lance},
                text width=68mm,
                align=left]  
            [Emotional prompting \cite{vinay2025emotional},
                text width=68mm,
                align=left
            ]
        ]
        [Content Camouflage and Evasion (\S \ref{sec:attack_content_3}),
            text width=54mm     
            [Detector evasion \cite{lularge2024,wu2024fake,tahmasebi2026robust,sakib2026credibility,wang2026prompt,das2025fake},
                text width=68mm,
                align=left] 
            [Jailbreak attack \cite{kaneko2026jailnewsbench},
                text width=68mm,
                align=left] 
            [Adversarial examples \cite{chen2023anti,deverna2024fact,danry2025deceptive,zhao2025lance,xu2025ssa,przybyla2025attacking,lu2026llm}, 
                text width=68mm,
                align=left
            ]
        ]
        [Scenario-specific Ecosystem Risks (\S \ref{sec:attack_content_4}),
            text width=54mm    
            [News recommendation pollution \cite{hu2025llm,zhao2025lance}, 
                text width=68mm,
                align=left
            ] 
            [Knowledge-intensive task pollution \cite{pan2023risk,pan2023attacking}  
             ,
                text width=68mm,
                align=left
            ]
        ]
    ]
    [Social-level (\S \ref{sec:attack_social}),
        text width=26mm
        [Comment Manipulation (\S \ref{sec:attack_social_1}),
            text width=54mm,
            align=left
            [Narrative-level amplification \cite{bandara2024hallucination,kim2025breaking,brian2025mpcg,liu2025stepwise},
                 text width=68mm,
                align=left
            ]
            [Message/comment generation \cite{luo2024message,le2020malcom,huynh2024improving,underwood2026generating,ahmed2026new}, 
                 text width=68mm,
                align=left
            ]
            [Real-comment retrieval attack \cite{koren2025evaluating},
                 text width=68mm,
                align=left
            ]
        ]
        [Structure Manipulation (\S \ref{sec:attack_social_2}),
             text width=54mm
            [User engagement manipulation \cite{wang2023attacking,wang2024bots},
                 text width=68mm,
                align=left
            ]
            [Graph structure attack \cite{zhu2024general} 
            ,
                 text width=68mm,
                align=left
            ]
        ]
        [Strategic Campaign Manipulation (\S \ref{sec:attack_social_3}),
             text width=54mm
            [RL-based comment/bot control \cite{le2022socialbots,yang2025robctrl,tong2025group},
                 text width=68mm,
                align=left
            ]
            [Campaign simulation \cite{qiao2025botsim,qiao-etal-2025-dynamic,lu2026large,jajanidze2025large},
                 text width=68mm,
                align=left
            ]
            [Actor identity manipulation \cite{feng2024does},          
                 text width=68mm,
                align=left
            ]
        ]
        [Propagation Data Poisoning(\S \ref{sec:attack_social_4}),
             text width=54mm
            [Backdoor attack \cite{jin2025backdoor,goschprovable},
                 text width=68mm,
                align=left
            ]
            [Platform data poisoning \cite{yamashita2024fake},
                 text width=68mm,
                align=left
            ]
        ]
    ]
    [Evidence-level (\S \ref{sec:attack_evidence}),
         text width=26mm
        [Evidence Content Manipulation (\S \ref{sec:attack_evidence_1}),
             text width=54mm
            [Synthetic Evidence Generation \cite{du2022synthetic,wan2025risk,abdelnabi2023fact,yang2026steering},  
                text width=68mm,
                align=left
            ]
            [Claim-driven Evidence Shift \cite{leite2026llm,layne2025analyzing,islam2025inconsistent,ou2026deceive,bethany2025camouflage},   
                text width=68mm,
                align=left
            ]
        ]
        [Evidence Corpus Poisoning (\S \ref{sec:attack_evidence_2}),
             text width=54mm
            [Targeted Knowledge Poisoning \cite{zou2025poisonedrag,wu2025admit,zhang2026practical,he2026fact2fiction,chang-etal-2025-one} \\ \cite{zhong2023poisoning,li2025cpa,chen2025mirage,geng2025unic,chen2025poisonarena,gong2025topic,chen2025flippedrag,wang2026joint},   
                text width=68mm,
                align=left
            ]
            [Structured Knowledge Poisoning \cite{chen2026kepo,cheng2024trojanrag,liang2025graphrag},  
                text width=68mm,
                align=left
            ]
            [Reasoning-pattern Knowledge Poisoning \cite{song2025chain,song2026adversarialcot},
                 text width=68mm,
                align=left
            ]
        ]
        [Evidence Retrieval Manipulation (\S \ref{sec:attack_evidence_3}),
             text width=54mm
            [Retrieval Visibility Enhancement \cite{cho2024typos,song2025silent,chang2026overcoming,nestaas2025adversarial},   
                text width=68mm,
                align=left
            ]
            [Retrieval Suppression \cite{shafran2025machine},  
                text width=68mm,
                align=left
            ]
        ]
    ]
]
\end{forest}
\end{adjustbox}
\caption{The taxonomy of LLM-enabled misinformation attacks.}
\label{fig:taxonomy}
\end{figure*}

LLMs introduce new risks to the misinformation ecosystem not simply because they can generate false text, but because their core capabilities can be misused as flexible attack engines. Natural-language generation supports low-cost fabrication and rewriting; contextual adaptation enables localized and audience-specific persuasion; persona simulation and interaction support social manipulation. These capabilities amplify the scale, automation, adaptivity, and accessibility of traditional misinformation campaigns, making content, social contexts, and evidence environments easier to manipulate.

In this section, we summarize LLM-empowered misinformation attacks from the attacker perspective. 
We organize the discussion around three manipulation surfaces: content-level, social-level, and evidence-level threats.
As shown in Fig.~\ref{fig:attack_overview}, content-level attacks manipulate the misinformation artifact itself, including its wording, style, framing, and detectability. 
Social-level attacks manipulate the surrounding social context, such as comments, personas, engagements, bot behaviors, and propagation traces. 
Evidence-level attacks manipulate the external evidence environment on which fact-checking, retrieval-augmented generation, and verification systems rely.

\subsection{Content-level Threats}
\label{sec:attack_content}
Content-level threats refer to the misuse of LLMs to generate or manipulate misleading information artifacts, such as fake news articles, social media posts, misleading answers, rewritten claims, and adversarially crafted texts. 
This line of research focuses on how false or misleading content is generated, linguistically adapted, and rhetorically framed. 
Existing studies can be grouped into the following four categories.

\subsubsection{Synthetic misinformation generation} \label{sec:attack_content_1}
Synthetic misinformation generation is the most direct form of LLM misuse studied in the literature. 
Many works study how adversaries prompt general-purpose LLMs to generate fluent, coherent, and news-like misinformation with minimal human effort \cite{vykopal2024disinformation,lucas2023fighting,sallami2024deception,wang2025have,pan2023risk,pan2023attacking,su2023fake,chen2024can}. 
These studies show that LLM-generated misinformation can resemble human-written content and can be adapted across topics, events, languages, and local contexts. 
Recent work further investigates scalable generation pipelines \cite{puccetti2024ai} and retrieval-augmented generation for producing more convincing or knowledge-grounded fake news \cite{singh2024adversarial}.

\subsubsection{Audience-aware persuasion} \label{sec:attack_content_2}
Another line of work studies audience-conditioned misinformation generation. 
Instead of generating a single generic false claim, these studies focus on how misleading narratives can be personalized according to user attributes, community contexts, ideological cues, or emotional states \cite{zugecova2025evaluation,zhao2025lance}. 
Related work on emotional prompting further investigates how affective framing, such as fear, anger, urgency, or moral concern, can increase the salience and persuasive force of generated misinformation \cite{vinay2025emotional}. 
This category reflects a shift from generic fake-content generation to targeted and context-aware influence.

\subsubsection{Content camouflage and defense evasion} \label{sec:attack_content_3}
A third group of studies investigates how misleading content can be rewritten to evade detection or moderation. 
These attacks preserve the deceptive meaning of a claim while changing surface style, sentiment, credibility cues, or linguistic form, making the content less recognizable to detectors that rely on textual artifacts or distributional patterns \cite{lularge2024,wu2024fake,tahmasebi2026robust,sakib2026credibility,wang2026prompt,das2025fake}. 
Recent studies also investigate jailbreak-based fake news generation, showing that harmful misinformation can still be elicited despite the presence of model safeguards \cite{kaneko2026jailnewsbench}. 
In addition, adversarial example generation uses explanation-guided rewriting, semantic contamination, iterative mutation, or fact-checking feedback to craft texts against fake news, rumor, or fact-checking systems \cite{lu2026llm,danry2025deceptive,deverna2024fact,zhao2025lance,xu2025ssa,chen2023anti}. 
These studies show that content-level attacks increasingly separate deceptive semantics from detectable surface cues.

\subsubsection{Application-oriented content attacks} \label{sec:attack_content_4}
Some studies examine how LLM-generated or rewritten misinformation affects downstream information systems. 
In news recommendation, manipulated content may alter ranking behavior, exposure patterns, and the visibility of reliable information \cite{hu2025llm,zhao2025lance}. 
In knowledge-intensive applications such as open-domain question answering, generated misinformation may appear as fluent answers or knowledge-like content, thereby polluting the information environment used by end users \cite{pan2023risk,pan2023attacking}.

\subsection{Social-level Threats} \label{sec:attack_social}
Social-level threats refer to LLM misuse that targets the social context in which misinformation propagates. 
Unlike content-level attacks that manipulate the claim or article itself, social-level attacks manipulate the social propagation process through which information spreads such as comments, user engagements, user profiles and bot behaviors. 
Existing studies can be grouped into the following four categories.

\subsubsection{Comment manipulation}
\label{sec:attack_social_1}
Comment manipulation studies focus on adversarial comments, replies, or conversational traces around a news item. 
Some studies explore narrative-level amplification, where LLM agents or persona-conditioned generation iteratively reshape misleading claims to make them more persuasive, coherent, or ideologically aligned during propagation \cite{bandara2024hallucination,kim2025breaking,brian2025mpcg,liu2025stepwise}. 
Some works focus on message/comment generation attacks, where adversarial textual signals are injected into the social context to mislead detectors \cite{luo2024message,huynh2024improving,le2020malcom,tong2025group,underwood2026generating,ahmed2026new}. 
Retrieval-based attacks further show that selectively inserted real user comments can also serve as adversarial payloads, even without generating new comments \cite{koren2025evaluating}. 

\subsubsection{Structure manipulation}
\label{sec:attack_social_2}
Structure manipulation studies examine attacks on the interaction patterns through which misinformation appears to spread. 
User-engagement manipulation perturbs sharing, reposting, or bot-generated interaction signals to make fake news appear more credible, popular, or organically endorsed \cite{wang2023attacking,wang2024bots}. 
Graph-structure attacks further manipulate the social interaction graphs used by graph-based fake news detectors, exposing the vulnerability of GNN-based models to propagation-level manipulation \cite{zhu2024general}.

\subsubsection{Strategic campaign manipulation}
\label{sec:attack_social_3}
Some studies treat propagation manipulation as a long-horizon campaign rather than a one-shot perturbation. 
RL-based social bot control formulates adversarial bot behavior as a sequential decision-making problem, where bots learn when to interact and evade detection to maximize influence over time \cite{le2022socialbots,yang2025robctrl}. 
LLM-powered campaign simulation further studies how malicious agents or botnets coordinate posting, commenting, and correction dynamics across multiple rounds of social interaction \cite{qiao2025botsim,qiao-etal-2025-dynamic,lu2026large}. 
\citet{jajanidze2025large} show the abuse of LLMs can also facilitate high-level social engineering by producing personalized behaviors.
\citet{feng2024does} examine how account attributes can be modified to make automated actors appear more human-like and persistent in online communities. 
These studies suggest that social-level attacks are moving toward a strategic campaign-level threat rather than an isolated perturbation.

\subsubsection{Propagation data poisoning} \label{sec:attack_social_4}
Propagation manipulation can also occur during data collection or model training. 
Backdoor attacks implant trigger patterns into propagation structures so that rumor detectors fail under specific conditions \cite{jin2025backdoor}. 
Robustness studies examine whether GNN-based detectors can be made provably robust against poisoning and backdoor attacks \cite{goschprovable}. 
Platform-level data poisoning further shows that fabricated user-generated content can contaminate the input distribution of online platforms and downstream detection systems \cite{yamashita2024fake}. 
Collectively, these studies highlight that propagation-based models are vulnerable not only at inference time, but also during the construction of the social data on which they are trained.

\subsection{Evidence-level Threats}
\label{sec:attack_evidence}

Evidence-level threats refer to attacks on the external evidence environment. 
These attacks manipulate what evidence is created, stored, retrieved, ranked, or used before a verification system reaches a judgment. 
Existing studies can be grouped into the following three categories.

\subsubsection{Evidence content manipulation}
\label{sec:attack_evidence_1}
Evidence content manipulation studies how misleading or adversarial evidence can alter verification. 
Some studies focus on synthetic evidence generation, adversarial document insertion, or evidence rewriting, where biased claims are made to appear supported \cite{du2022synthetic,wan2025risk,abdelnabi2023fact}. 
More recent work extends this to viewpoint steering, where fabricated rationales guide models toward attacker-preferred perspectives \cite{yang2026steering}. 
Another line of research manipulates the input claim rather than the evidence source itself. 
They change search queries to manipulate the evidence matching and verification trajectories via adversarially paraphrasing \cite{layne2025analyzing,bethany2025camouflage,islam2025inconsistent,leite2026llm,ou2026deceive}. 
These studies show that evidence-level failures can arise either from evidence content or from redirecting the evidence acquisition and reasoning processes via claim reformulation.

\subsubsection{Evidence corpus poisoning}
\label{sec:attack_evidence_2}

Evidence corpus poisoning studies examine attacks that inject malicious documents into retrieval corpora or knowledge bases used by RAG and fact-checking systems. 
Targeted poisoning methods craft passages for specific queries, claims, entities, or verdicts, so that the system retrieves poisoned evidence and produces attacker-desired answers or fact-checking decisions \cite{zhong2023poisoning,zou2025poisonedrag,wu2025admit,zhang2026practical,chang-etal-2025-one,he2026fact2fiction,wang2026joint}. 
Some studies further broaden this setting to black-box, transferable, competing, or stealthy poisoning, showing that poisoned evidence may remain effective across different queries, topics, or detection constraints \cite{chen2025mirage,geng2025unic,chen2025poisonarena,li2025cpa}. 
More recent works extend poisoning objectives beyond isolated factual answers, including viewpoint manipulation, trigger-conditioned evidence stores, and reasoning-path corruption in retrieved documents \cite{gong2025topic,chen2025flippedrag,chen2026kepo,cheng2024trojanrag,liang2025graphrag,song2025chain,song2026adversarialcot}. 
These studies show that corpus poisoning can affect not only what knowledge is retrieved, but also how the model reasons over retrieved evidence.

\subsubsection{Evidence retrieval manipulation}
\label{sec:attack_evidence_3}

Evidence retrieval manipulation studies target the retrieval stage itself. 
One group of attacks increases the visibility of malicious evidence by crafting or perturbing documents so that adversarial content is more likely to enter the retrieved context \cite{chang2026overcoming,nestaas2025adversarial}. 
Related methods use typos, mutations, or imperceptible textual changes to alter retrieval behavior while preserving the apparent meaning of the document \cite{cho2024typos,song2025silent}. 
Another group of attacks suppresses useful evidence by introducing jamming documents that interfere with retrieval-augmented generation and make correct supporting evidence harder to access or use \cite{shafran2025machine}. 
These studies show that evidence integrity depends not only on whether malicious evidence exists, but also on whether reliable evidence remains retrievable, salient, and properly ranked.

\subsection{Summary}
Overall, existing research has shown that the misuse of LLMs enables misinformation risks from isolated content fabrication into ecosystem-level manipulation. 
\begin{itemize}
\item[-] Content-level threats have evolved from simple fake-text generation to adaptive misinformation production. 
Early attacks primarily use LLMs to produce fluent fake posts, whereas recent studies increasingly exploit LLMs for personalization, emotional framing, multilingual rewriting, detector evasion, and downstream application attacks. 
The key trend is that content-level attackers become more scalable, audience-aware, and defense-aware. 

\item[-] Social-level threats have evolved from isolated comment manipulation to coordinated propagation manipulation. 
Early attacks primarily inject adversarial comments or reactions to mislead detectors, whereas recent work increasingly manipulates engagement structures, bot behaviors, user identities, and long-horizon campaign dynamics. 
The key trend is that attackers no longer only manipulate what users read, but also how socially credible, controversial, or organically endorsed the information appears. 

\item[-] Evidence-level threats have evolved from attacking model outputs to compromising the evidential infrastructure that verification systems rely on. Early attacks primarily fabricate or rewrite supporting evidence, whereas recent work increasingly studies corpus poisoning, retrieval manipulation, viewpoint steering, universal poisoning, stealthy poisoned documents, and reasoning-path corruption. 
The key trend is that attackers can compromise not only what evidence exists, but also what evidence is retrieved, and used in reasoning. 
\end{itemize}

The three attack layers differ in both threat assumptions and evaluation requirements. 
Content-level attacks assume control over the misinformation artifact itself and are typically evaluated by generation quality, semantic preservation, persuasive effect, or detector evasion. 
Social-level attacks manipulate interaction contexts such as comments, user profiles, engagement traces, or bot behaviors, thereby changing how credible, controversial, or organic a claim appears. 
Evidence-level attacks target verification systems that rely on external evidence, retrieval corpora, search engines, or knowledge bases, manipulating what evidence is created, retrieved, ranked, and used in reasoning. 
These differences also lead to different data needs: content attacks can often be evaluated on text or multimodal misinformation datasets, whereas social- and evidence-level attacks require propagation traces, campaign simulations, claim-evidence corpora, or RAG-style benchmarks.

\section{LLMs as Victims: Vulnerabilities of LLM-based Detection Paradigms} \label{sec:victims} 

\begin{figure}[t]
    \centering
    \includegraphics[width=0.7\linewidth]{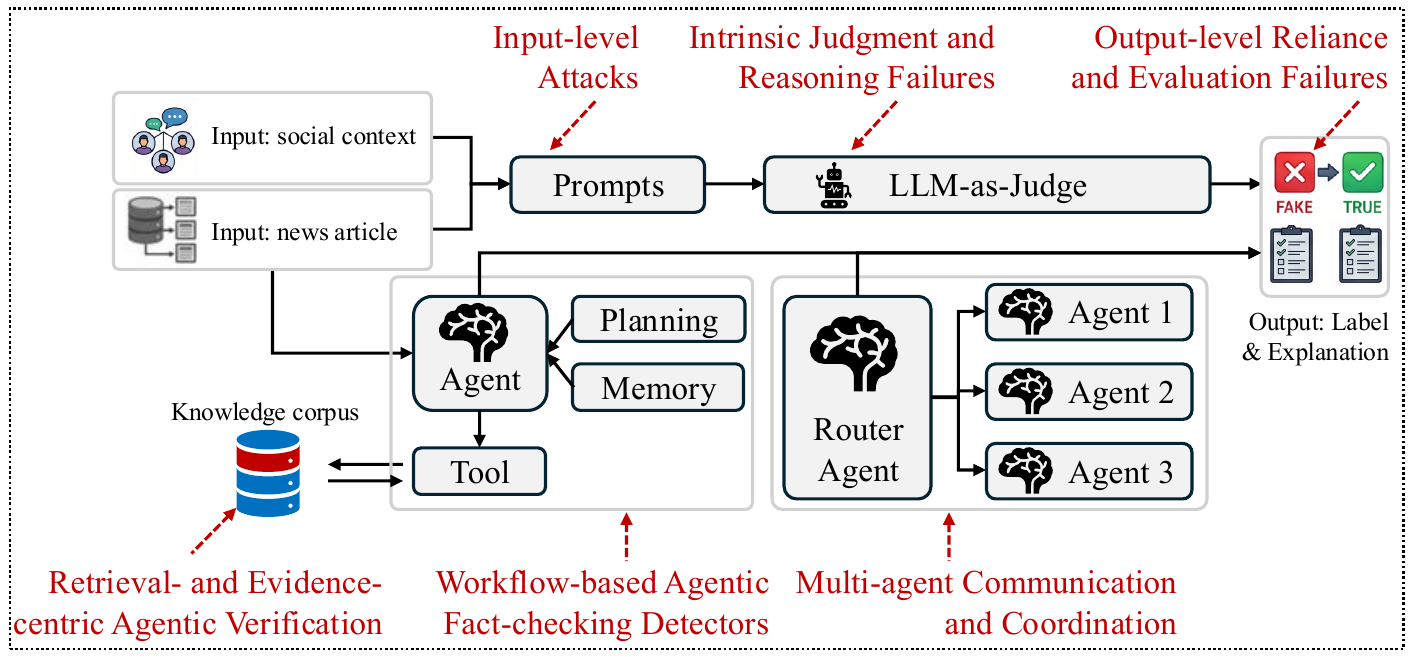}
    \caption{Overview of vulnerabilities of LLM-based misinformation detection paradigms.}
    \label{fig:victim_overview}
\end{figure}

\begin{figure*}[t]
\centering
\small
\begin{adjustbox}{max width=0.9\textwidth, max height=0.88\textheight, center}
\begin{forest}
for tree={
    grow'=east,
    draw,
    rounded corners=3pt,
    align=left,
    font=\small,
    inner sep=3pt,
    edge={draw, -},
    parent anchor=east,
    child anchor=west,
    l sep=12mm,
    s sep=3mm,
    forked edges,
    fork sep=2mm,
    execute at begin node=\RaggedRight\sloppy,   
}
[\rotatebox{0}{LLM as Victim (\S \ref{sec:victims})}, 
    fill=white,
    [LLM-based Detector (\S \ref{sec:victim_1}),
        text width=40mm
        [Intrinsic judgment and reasoning failures 
        \cite{chen2024can,ma2025linguistic,qi2025evaluating,wang2025towards,peng2024securing}, 
            text width=100mm
        ]
        [Input-level attacks on LLM judges \cite{maloyan2025adversarial,sili2025universal,raina2024llm,shi2024optimization,przybyla2025verifying,leite2026llm},
            text width=100mm
        ]
        [Output-level reliance and evaluation failures \cite{deverna2024fact,schwinn2026coin,shafee2026false}, 
            text width=100mm
        ]
    ]
    [Agentic Detector System (\S \ref{sec:victim_2}),
        text width=40mm
        [Workflow-based agentic fact-checking detectors \cite{li2024large_ecai,li2024large,li2026factguard,huang2026diva,ahmad2025urdufactcheck}, 
            text width=100mm
        ]
        [Retrieval- and evidence-centric agentic verification \cite{tan2024glue,xue2024badrag,he2026fact2fiction,chen2025mirage,wu2025exclaim,liu2026raar}, 
            text width=100mm
        ]
        [Multi-agent communication and coordination \cite{aldahoul2025toward,lee2025prompt,muneer2026mosaiv,bukke2025agentic,avram2025mcp}, 
            text width=100mm
        ]
    ]
]
\end{forest}
\end{adjustbox}
\caption{The taxonomy of vulnerabilities of LLM-based misinformation detection paradigms.}
\label{fig:taxonomy}
\end{figure*}
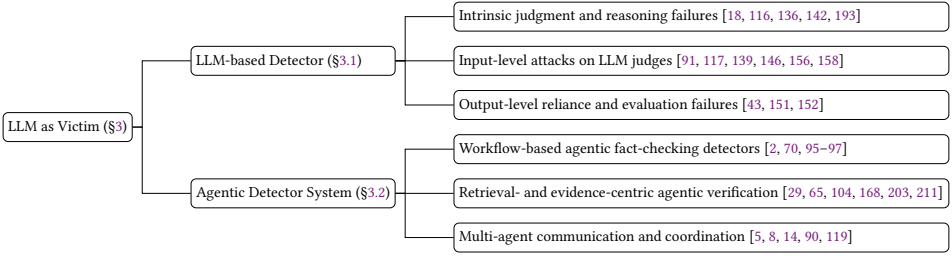

In this section, we study LLMs as attack targets within misinformation detection systems that either directly employ LLMs as judges or organize them into agentic workflows for misinformation detection.
This section differs from Sec.~\ref{sec:defender} in perspective, which aims to review how LLM-based systems are designed to detect and mitigate misinformation. In contrast, this section examines how the LLM-centric systems may fail when they are attacked adversarially. 
We organize this section by the primary attack surface of LLM-based misinformation detection systems. 
As shown in Fig.~\ref{fig:victim_overview},
for direct LLM-based detectors, the attack surface follows an input/judgment/output structure. 
Adversaries may manipulate the input, exploit the model's internal judgment weaknesses, or distort downstream decisions through unreliable outputs. 
For agentic detectors, the attack surface expands from a single model call to a broader workflow involving planning, retrieval, tools, communication, memory, trust, and aggregation.

\subsection{LLM-as-judge Vulnerabilities}
\label{sec:victim_1}
Direct LLM-based detectors use an LLM as a judge or verifier. 
Given a claim, the LLM is prompted to produce a veracity label, factuality judgment, or hallucination assessment. 
This paradigm is attractive because it requires limited task-specific architectural design and can be adapted to different misinformation-related tasks through prompting. 
However, it also inherits the vulnerabilities of LLM-as-a-judge systems, particularly when exposed to adversarially crafted inputs, ambiguous factual contexts, or excessive downstream reliance on the LLM's judgments.

\subsubsection{Input-level Attacks on LLM Judges}
\label{subsubsec:input_level_attacks_llm_judges}

The first vulnerability arises from the input interface of LLM-based detectors. 
Because task instructions and untrusted content are processed in the same natural-language context, adversaries can manipulate the detector before factual judgment begins. 
Several studies directly expose this input-level vulnerability. 
Some attacks directly inject malicious instructions or optimized prompt sequences into attacker-controlled inputs, thereby biasing LLM-as-a-judge systems before they perform factual assessment \citep{maloyan2025adversarial,shi2024optimization,sili2025universal}. 
Others rely on short transferable adversarial phrases that can distort zero-shot LLM assessments across different judge settings \citep{raina2024llm}. 
The manipulation can also be more semantic and less visible via meaning-preserving perturbations \cite{przybyla2025verifying} or adversarial persuasive attacks  \cite{leite2026llm}. 

\subsubsection{Intrinsic Judgment and Reasoning Failures}
\label{subsubsec:intrinsic_judgment_reasoning_failures}
The second vulnerability stems from the LLM's own judgment and reasoning behavior. 
Even in the absence of explicit prompt injection, a direct LLM-based detector may fail because it relies on linguistic cues, rhetorical fluency, or plausible yet invalid reasoning chains. 
In misinformation detection, this is especially problematic because false claims are often written in polished language and may be supported by fabricated explanations that appear logically coherent.
Existing studies suggest that this vulnerability arises from several sources. 
First, LLM-based detectors may fail when synthetic misinformation becomes fluent, human-like, and difficult to distinguish from authentic content \citep{chen2024can,ma2025linguistic}. 
This indicates that direct detectors cannot safely rely on surface-level artifacts or stylistic cues as LLM-generated misinformation becomes more natural and adaptive.
Second, LLM-based detectors may be unstable when factual judgment requires resolving conflicts with the model's parametric knowledge \citep{qi2025evaluating}. 
Third, LLMs may be biased toward reasoning that appears coherent but is logically or factually invalid \citep{wang2025towards}, which can cause rationale-based detectors to overvalue fluent explanations instead of verifying their factual grounding. 
Broader security analyses further suggest that bias, misinformation, and prompt-based manipulation are intertwined risks in LLM-based applications \citep{peng2024securing}.

\subsubsection{Output-level Reliance and Evaluation Failures}
\label{subsubsec:output_reliance_evaluation_failures}

The third vulnerability arises after the LLM detector produces its output. 
LLM-based detectors often generate not only a binary label, but also explanations, confidence estimates, factuality rationales, or safety judgments. 
These outputs can influence human users, content moderators, evaluation pipelines, and platform-level decision systems. 
Consequently, the harm of a detector is not limited to whether its label is correct; it also depends on how its output is interpreted and used downstream.
Output-level failures can affect both users and evaluation pipelines. 
\citet{deverna2024fact} show that fact-checking information generated by large language models can decrease headline discernment under certain conditions. 
\citet{schwinn2026coin} argue that attack success may sometimes reflect weaknesses of the judge rather than genuine failures of the target system. 
\citet{shafee2026false} show that adversarial attacks against LLM-based cyber threat intelligence systems can cause false alarms and real downstream damage.

\subsection{Vulnerabilities of Agentic Misinformation Detection Systems}
\label{sec:victim_2}

Agentic misinformation detectors differ from direct LLM-based detectors because they do not simply ask an LLM for a one-shot judgment. 
Instead, they organize LLMs into multi-step workflows that may decompose claims, retrieve evidence, invoke tools, coordinate multiple agents, store context, and aggregate intermediate results. 
This paradigm can emulate human fact-checking more closely and can improve transparency, adaptability, and evidence grounding. 
However, the same workflow structure also expands the attack surface. 
An adversary can target not only the final LLM judge, but also the planning process, evidence environment, tool chain, inter-agent communication, memory, trust estimation, or decision policy.

\subsubsection{Workflow-based Agentic Fact-checking Detectors}
\label{subsubsec:workflow_agentic_factchecking}
This category includes agentic detectors whose primary contribution is the construction of a structured verification workflow. 
Their central idea is to move beyond one-shot LLM classification by approximating the behavior of human fact-checkers through claim decomposition, stepwise reasoning, tool selection, and final verdict integration \cite{li2024large_ecai,li2026factguard,huang2026diva,ahmad2025urdufactcheck}. 
The vulnerability of this category lies in the workflow itself. 
Many works on general agentic systems show that autonomous LLM agents may suffer from unsafe planning, tool misuse, context manipulation, runtime supply-chain attacks, and uncontrolled interactions \cite{deng2025ai,chhabra2026agentic,jiang2026agentic}. 
These risks are especially relevant to misinformation detection because agentic fact-checking systems typically rely on web search, external databases, browser tools, and other runtime resources that can themselves be manipulated. 
If the system decomposes a claim incorrectly, selects the wrong tool, overweights a misleading subclaim, or follows an unsafe reasoning path, the final verdict may fail even if the base LLM is capable. 

\subsubsection{Retrieval- and Evidence-centric Agentic Verification}
\label{subsubsec:retrieval_evidence_agentic_verification}

The category includes agentic systems whose main contribution centers on external evidence. 
Several agentic detectors rely on retrieval and web grounding \cite{tan2024glue,xue2024badrag,chen2025mirage,wu2025exclaim,liu2026raar}. 
External evidence can reduce hallucination and improve grounding, but it can also be poisoned, fabricated, selectively retrieved, or adversarially ranked. 
Fact2Fiction \cite{he2026fact2fiction} targets agentic fact-checking systems by constructing poisoned evidence tailored to the system's claim decomposition strategy. 
This demonstrates that evidence-grounded detectors can be compromised through the evidence ecosystem even when the original claim and the base LLM remain unchanged.
\citet{singh2026adversarial} extend general security framing to  multimodal agentic RAG. 
Its central idea is that adversarial intent may be distributed across retrieval, planning, and generation stages rather than appearing as a single malicious input. 
This motivates stateful trust modeling, where the system tracks potentially adversarial behavior across the workflow.  

\subsubsection{Multi-agent Communication and Coordination}
\label{subsubsec:multiagent_communication_coordination}

The category includes systems whose main contribution or vulnerability comes from multiple interacting agents. 
Multi-agent misinformation detectors often assign different roles to specialized agents, such as claim analysis, evidence retrieval, debate, and final judgment \cite{aldahoul2025toward,muneer2026mosaiv,bukke2025agentic,avram2025mcp}. 
This design can improve coverage and interpretability, but it also introduces communication channels through which errors, biased intermediate conclusions, or malicious prompts can propagate. 
The distinctive vulnerability of this category is not merely that each agent may be wrong, but that the system depends on communication and coordination among agents. 
\citet{lee2025prompt} reveal LLM-to-LLM prompt injection within multi-agent systems, showing that malicious prompts can spread across interconnected agents.

\subsection{Summary}
\label{subsec:victim_discussion}
This section shows that vulnerabilities of LLM-based systems evolve with the detection paradigm. 
\begin{itemize}
\item[-] For direct LLM-based detectors, robustness centers on the input, the model's internal judgment, and the downstream interpretation of its output. 
An adversary can inject malicious instructions, craft persuasive false claims, exploit reasoning bias, or cause users and evaluation systems to overtrust unreliable explanations. 
\item [-] For agentic detectors, robustness becomes a workflow-level problem. 
The adversary can target the detector's planning process, evidence environment, runtime tools, communication channels, stateful trust mechanisms, or coordination policies.
\end{itemize}
These observations imply that improving the underlying LLM alone is insufficient. 
Direct LLM-based detectors require prompt isolation, adversarial input testing, and uncertainty calibration.
Agentic detectors require provenance-aware retrieval, secure tool use, authenticated inter-agent communication, memory auditing, trust modeling, and robust coordination policies. 
As misinformation detection systems become increasingly agentic, robustness evaluation must move from isolated model accuracy toward end-to-end security analysis of the entire verification workflow.

\section{LLMs as Defenders: LLM-enabled Detection and Verification Methods}\label{sec:defender}

\begin{figure}[t]
    \centering
    \includegraphics[width=0.7\linewidth]{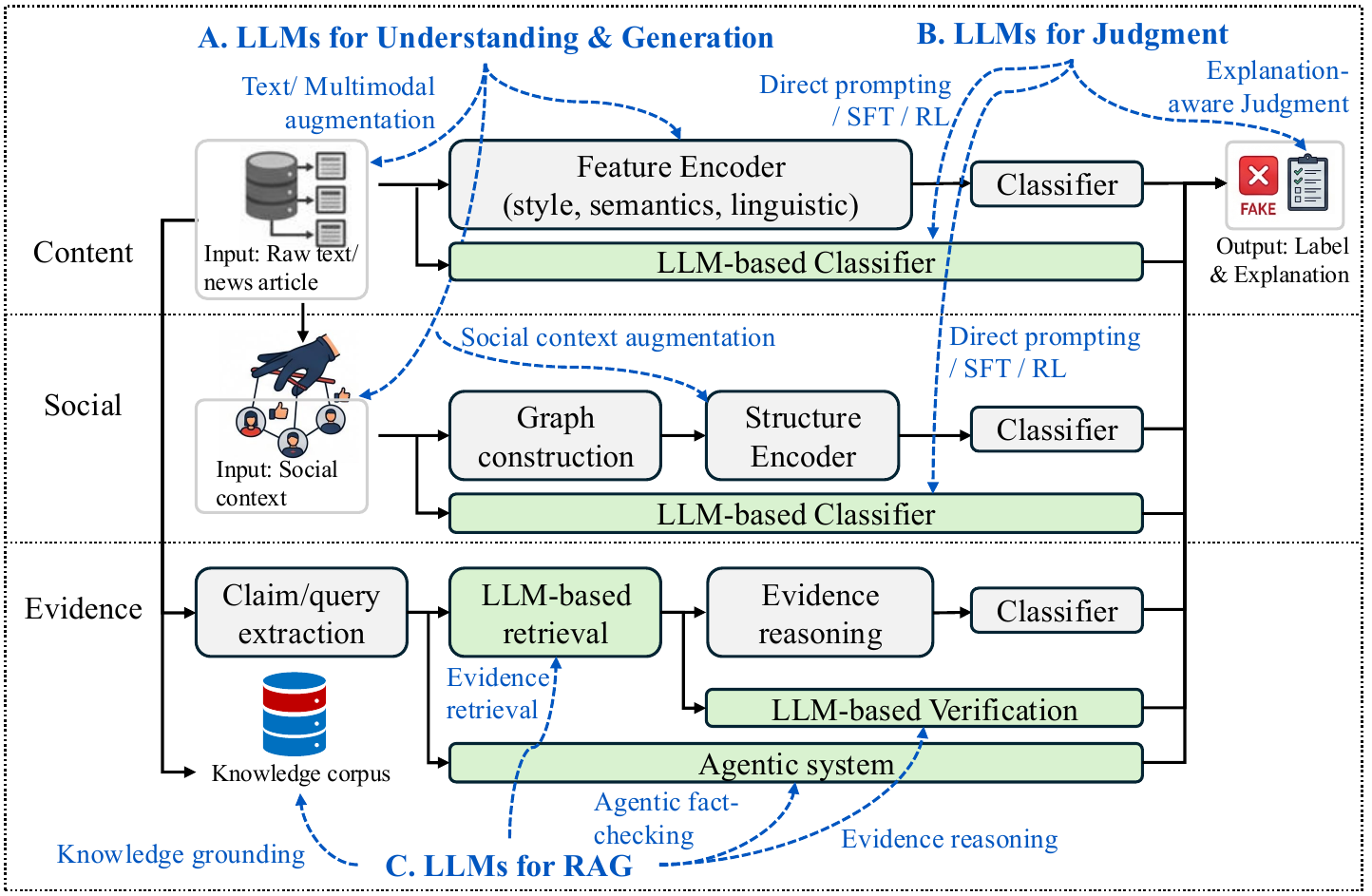}
    \caption{Overview of LLM-based defense methods for misinformation detection.}
    \label{fig:defense_overview}
\end{figure}

\begin{figure*}[!ht]
\centering
\begin{adjustbox}{max width=\textwidth, center}
\small
\begin{forest}
for tree={
    grow'=east,
    draw,
    rounded corners=3pt,
    align=left,
    font=\small,
    inner sep=3pt,
    edge={draw, -},
    parent anchor=east,
    child anchor=west,
    l sep=8mm,
    s sep=3mm,
    forked edges,
    fork sep=2mm,
    execute at begin node=\RaggedRight\sloppy,
}
[\rotatebox{90}{LLMs as Defenders (\S \ref{sec:defender})},  
    [Understanding \\ \& Generation \\ (\S \ref{subsec:understanding_generation}),
         text width=23mm
        [Text Augmentation (\S \ref{subsubsec:text_augmentation}),  
             text width=50mm
            [Style-aware augmentation \cite{park2025adversarial,shen2025llm,zhou2026latent},  
                text width=70mm
            ]
            [Semantic augmentation \cite{wu2024towards,tian2025symbolic,DBLP:conf/aaai/CaoWZH25}, 
                text width=70mm
            ] 
        ]
        [Social Context Augmentation (\S \ref{subsubsec:social_context_augmentation}), 
             text width=50mm
            [Propagation modeling \cite{chen2025explore,zeng2025exploring,jiang2025epidemiology,zeng2026human}, 
                text width=70mm
            ]
            [Propagation data simulation \cite{nan2024let,chen2025structure,chen2026information},  ,
                text width=70mm
            ]
        ]
        [Multimodal Augmentation (\S \ref{subsubsec:multimodal_augmentation}),
             text width=50mm
            [Multimodal reasoning \cite{habib2026llm,yan2025collaborate,zheng2025unveiling},  
                text width=70mm
            ]
            [Missing Modality Simulation \cite{qian2026head},  
                text width=70mm
            ]
        ]
    ]
    [Judgment (\S \ref{subsec:judgement}),
        text width=23mm
        [Direct Prompting (\S \ref{subsubsec:direct_prompting}),  
             text width=50mm
            [Prompting \cite{turaga2024information,su2024adapting,chen2024can,chen2025explore},  
                text width=70mm
            ]
            [Chain-of-Thought Prompting \cite{xu2024multimodal,kareem2023fighting},                 text width=70mm
            ]
            [Role-playing \cite{hu2024bad}, 
                text width=70mm
            ]
        ]
        [Supervised Fine-tuning (\S \ref{subsubsec:supervised_tuning}),
             text width=50mm
            [Task-specific Fine-tuning \cite{yan2025collaborate,gong2025cross,zhou2025collaborative,dong2026dpsa},  
                 text width=70mm
            ]
            [Explanation-supervised Fine-tuning \cite{zheng2025predictions,wang2025llm,wang2026rationales},  
                text width=70mm
            ]
        ]
        [Reinforcement Learning-based Tuning \\ (\S \ref{subsubsec:reinforcement_tuning}), 
            text width=50mm
            [Reinforcement Learning  \cite{ding2025dynamic,yang2024reinforcement},  
                text width=70mm
            ]
            [Reinforcement-guided Sampling \cite{tong2025generate}, 
                text width=70mm
            ]
        ]
        [Explanation-aware Judgment (\S \ref{subsubsec:explanation_aware_judgement}),  
            text width=50mm
            [Multimodal Explanation \cite{zhou2026graph,qi2024sniffer},  
                text width=70mm
            ]
            [Debunking Explanation \cite{zeng2026manipulation}, 
                text width=70mm
            ]
        ]
    ]
    [RAG System (\S \ref{subsec:rag_system}),
        text width=23mm
        [Knowledge Grounding (\S \ref{subsubsec:knowledge_grounding}),  
            text width=50mm
            [Knowledge-aware Verification \cite{DBLP:journals/eaai/YanZW24,DBLP:conf/nlpcc/ChenW24}, 
                text width=70mm
            ]
            [Dynamic Knowledge Updating \cite{DBLP:conf/ijcai/0001YWZLH25}, 
                text width=70mm
            ]
        ]
        [Evidence Retrieval (\S \ref{subsubsec:evidence_retrieval}), 
            text width=50mm
            [Textual evidence retrieval \cite{ma2026radar,DBLP:conf/bigdataconf/BaiF24},  
                text width=70mm
            ]
            [Multimodal evidence retrieval \cite{DBLP:journals/information/DongCKWDL26,DBLP:journals/ipm/HanZWLD26},  
                text width=70mm
            ]
        ]
        [Evidence Reasoning (\S \ref{subsubsec:evidence_reasoning}),  
            text width=50mm
            [Semantic/Commonsense evidence reasoning \cite{wu2024towards,he2026factguard},  
                text width=70mm
            ]
            [Multimodal evidence reasoning \cite{zhou2026graph,qi2024sniffer},  
                text width=70mm
            ]
            [Trustworthy evidence reasoning \cite{wan2026facade,zhang2024causal},   
                text width=70mm
            ]
        ]
        [Agentic Fact-checking (\S \ref{subsubsec:agentic_fact_checking}), 
            text width=50mm 
            [Multi-agent System \cite{tyagi2025mavs,aldahoul2025toward,liu2025truth,DBLP:conf/mcsoc/DaoHT25,li2024large_ecai,hong2025emulate}, 
                text width=70mm
            ]
            [Debate-driven Verification \cite{wang2024explainable,he2026debating,han2026beyond,han2025debate}, 
                text width=70mm
            ]
            [Agentic Debunking \cite{zeng2026manipulation}, 
                text width=70mm
            ]
        ]
    ]
]
\end{forest}
\end{adjustbox}
\caption{The taxonomy of LLM-based defense methods.}
\label{fig:llm_defenders_taxonomy}
\end{figure*}
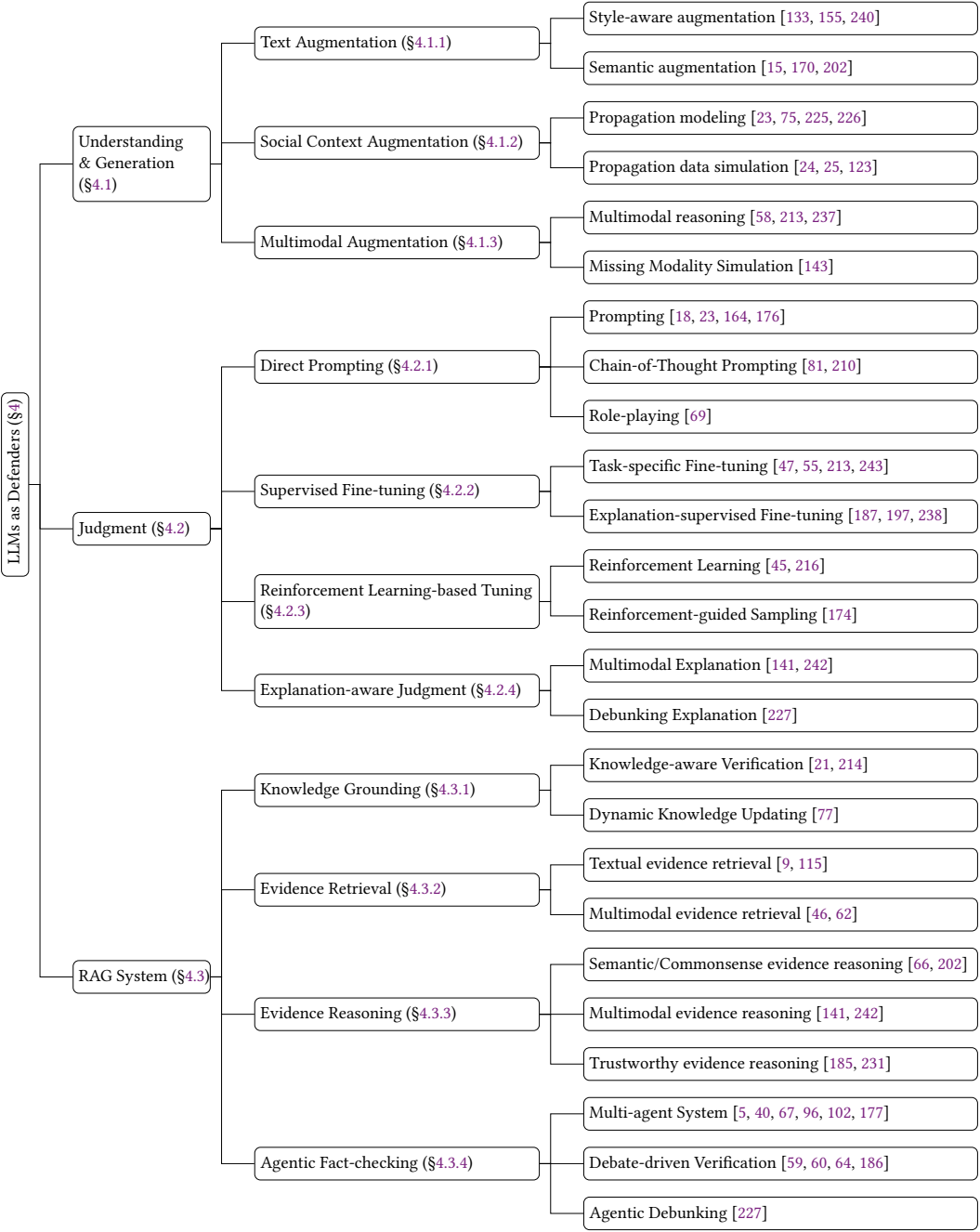

Large language models (LLMs) have been increasingly explored as defenders against misinformation.
As shown in Fig.~\ref{fig:llm_defenders_taxonomy}, we organize existing studies into three lines of work. 
The first uses LLMs as understanding and generation modules to enrich textual, social, and multimodal signals. 
The second uses LLMs as judgment modules through prompting, tuning, reinforcement learning, or explanation-aware prediction. 
The third integrates LLMs with retrieval-augmented generation (RAG), external evidence, and agentic workflows for more grounded verification.

\subsection{Understanding and Generation}
\label{subsec:understanding_generation}
The first line of research uses LLMs as auxiliary modules for data augmentation, representation enrichment, context simulation, and multimodal understanding. 
Existing studies can be grouped into text augmentation, social context augmentation, and multimodal augmentation.

\subsubsection{Text Augmentation}
\label{subsubsec:text_augmentation}

Text augmentation studies use LLMs to construct alternative textual views, semantic enrichments, or adversarial variants for misinformation detection. 
This is especially useful when detectors overfit to superficial writing styles or fail under LLM-enabled attacks.
Some methods reduce detectors' reliance on superficial writing patterns by exposing them to stylistic variants of the same misleading semantics \cite{park2025adversarial,shen2025llm,zhou2026latent}. 
Other works focus on enriching the latent representation via semantic relationship perception between content and supporting evidence \cite{wu2024towards}, or structured semantic signals \cite{tian2025symbolic,DBLP:conf/aaai/CaoWZH25}. 
These studies show that LLM-based augmentation can help detectors capture deeper semantic structures.

\subsubsection{Social Context Augmentation}
\label{subsubsec:social_context_augmentation}

Social context augmentation studies use LLMs to interpret, refine, or simulate the social signals surrounding misinformation. 
This line of research is motivated by the fact that comments, stance labels, propagation trees, and interaction patterns are often missing, noisy, or manipulated in real-world settings. 
Some methods use LLMs to interpret social reactions, infer user stances, and model propagation dynamics \cite{chen2025explore,zeng2025exploring,jiang2025epidemiology,zeng2026human}.
Other works simulate propagation data to address the lack of sufficient social feedback at the early stage or low-resource detection such as emerging topics \cite{nan2024let,chen2026information,chen2025structure}.

\subsubsection{Multimodal Augmentation}
\label{subsubsec:multimodal_augmentation}

Some studies extend LLM-based defense from text-only misinformation to settings involving visual evidence such as images and videos. 
The primary idea is to use LLMs or multimodal large language models (MLLMs) to enrich cross-modal representations, reason over text-image consistency, or handle missing modalities. 
Some methods use LLMs or MLLMs to jointly interpret textual and visual signals before making a detection decision \cite{habib2026llm,yan2025collaborate,zheng2025unveiling}. 
MLLMs also offer a new perspective to address the practical problem that misinformation samples may lack complete visual, textual, or metadata information \cite{qian2026head}.

\subsection{Judgment}
\label{subsec:judgement}
The second line of research directly uses LLMs as judgment modules. 
These studies use LLMs to generate veracity labels, reliability scores, and rationales. 
Compared with augmentation-based methods, judgment-based methods place LLMs closer to the final decision process. 
Existing studies mainly explore four technical ways: direct prompting, supervised tuning, reinforcement tuning, and explanation-aware judgment.

\subsubsection{Prompt-based Judgment}
\label{subsubsec:direct_prompting}
Some prompting-based methods reformulate misinformation detection as a natural-language judgment task via well-designed instructions \cite{turaga2024information,su2024adapting,chen2024can,chen2025explore}.
Some works further design chain-of-thought (CoT) prompts to decompose the LLM judgment process into intermediate reasoning steps \cite{xu2024multimodal,kareem2023fighting}. 
\citet{hu2024bad} explore the dual role of LLMs in fake news detection and show that LLMs may function as useful advisors even when they can also be misused as generators. 
They offer a low-cost entry point for LLM-based detection, but its reliability remains constrained by instruction sensitivity and unstable generalization.

\subsubsection{Supervised Fine-tuning}
\label{subsubsec:supervised_tuning}

Supervised fine-tuning methods adapt LLMs or LLM-enhanced detectors to misinformation-specific data, labels, domains, or rationales. 
Task-specific tuning is used for emergency rumors, cross-domain rumors, emergent fake news, and unseen misinformation, where models must adapt to evolving events and platforms \cite{yan2025collaborate,gong2025cross,zhou2025collaborative,dong2026dpsa}. 
Another line of research incorporates rationales or explanations into the training process, encouraging models to produce analytical outputs rather than only labels \cite{zheng2025predictions,wang2025llm,wang2026rationales}. 
These methods improve task alignment and stability compared with prompting, but they may introduce dataset-specific biases.

\subsubsection{Reinforcement Learning-based Tuning}
\label{subsubsec:reinforcement_tuning}
Reinforcement learning-based tuning methods optimize LLM-based detectors using reward signals, adversarial training, or reinforcement-guided sampling. 
Some studies formulate misinformation detection as an adaptive adversarial process, where detectors are trained against semantically constrained or dynamically generated fake information \cite{ding2025dynamic}. 
Others use reinforcement tuning to jointly detect stances and debunk rumors, connecting intermediate social reasoning with final veracity judgment \cite{yang2024reinforcement}. 
Reinforcement-guided sampling further combines LLM generation with feedback-based instance selection, enabling detectors to benefit from informative synthetic samples \cite{tong2025generate}. 
These methods are useful when misinformation detection involves evolving adversaries, multiple optimization objectives, or limited labeled data.

\subsubsection{Explanation-aware Judgment}
\label{subsubsec:explanation_aware_judgement}
Explanation-aware judgment methods treat explanations as part of the decision process. 
Some studies use graph-prompted or MLLM-based explanations to make decisions over text-image relations more interpretable \cite{zhou2026graph,qi2024sniffer}. 
\citet{zeng2026manipulation} move beyond label prediction toward debunking explanations, where the system identifies the misleading mechanism and communicates corrective information to users.  

\subsection{RAG System}
\label{subsec:rag_system}

The third line of research integrates LLMs with external knowledge, evidence retrieval, and agentic verification. 
Existing studies can be organized along the verification pipeline: knowledge grounding, evidence retrieval, evidence reasoning, and agentic fact-checking.

\subsubsection{Knowledge Grounding}
\label{subsubsec:knowledge_grounding}
Knowledge grounding methods enhance LLM-based verification by incorporating external or dynamically updated knowledge. 
Some studies use knowledge-guided prompting or retrieval-augmented LLMs to improve rumor detection and resolve unseen rumors when parametric knowledge is insufficient \cite{DBLP:journals/eaai/YanZW24,DBLP:conf/nlpcc/ChenW24}. 
Other work studies dynamic knowledge updating for fake news detection, aiming to align verification with evolving factual contexts around rapidly changing events \cite{DBLP:conf/ijcai/0001YWZLH25}. 

\subsubsection{Evidence Retrieval}
\label{subsubsec:evidence_retrieval}
Evidence retrieval methods acquire relevant textual, visual, or cross-modal evidence before verification. 
Text-based RAG studies retrieve documents, passages, or facts to support fake news detection and fact-checking decisions \cite{ma2026radar,DBLP:conf/bigdataconf/BaiF24}. 
Multimodal retrieval methods further combine external evidence with visual forensics or cross-modal retrieval for multimodal fake news and video misinformation detection \cite{DBLP:journals/information/DongCKWDL26,DBLP:journals/ipm/HanZWLD26}.  

\subsubsection{Evidence Reasoning}
\label{subsubsec:evidence_reasoning}
Evidence reasoning methods investigate how verification systems determine whether retrieved evidence supports, refutes, or is irrelevant to a claim. 
Some works enhance semantic understanding of claim-evidence relations or incorporate event-centric and commonsense-guided reasoning to move beyond surface matching \cite{wu2024towards,he2026factguard}. 
Graph-prompted and out-of-context detection methods reason over entities, captions, events, and visual scenes to identify cross-modal inconsistencies \cite{zhou2026graph,qi2024sniffer}. 
Another line of research focuses on trustworthy evidence reasoning, including LLM susceptibility to deceptive evidence and causal debiasing for multi-hop fact verification \cite{wan2026facade,zhang2024causal}.  

\subsubsection{Agentic Fact-checking}
\label{subsubsec:agentic_fact_checking}
Agentic fact-checking methods extend RAG-based verification from a single-step pipeline to an interactive, multi-step workflow. 
Multi-agent systems distribute verification across roles such as claim analyzer, evidence retriever, verifier, and explainer \cite{tyagi2025mavs,aldahoul2025toward,liu2025truth,DBLP:conf/mcsoc/DaoHT25,li2024large_ecai,hong2025emulate}. 
Debate-driven methods use disagreement and argumentation to expose weak evidence, unsupported assumptions, and one-sided reasoning, and have been studied for misinformation detection and intervention \cite{wang2024explainable,he2026debating,han2026beyond,han2025debate}. 
Agentic debunking further moves from veracity prediction to user-facing correction, requiring systems to identify misleading mechanisms, collect evidence, and generate corrective explanations \cite{zeng2026manipulation}.  

\subsection{Summary}
\label{subsec:defender_summary}
  
LLM-based misinformation defense is moving beyond conventional content classification toward more contextual, evidence-grounded, and workflow-oriented verification. 
\begin{itemize}
\item[-]  LLMs are mainly used to enrich the inputs and representations available to downstream detectors. Text augmentation improves robustness to stylistic variation and adversarial rewriting; social context augmentation helps recover or refine missing propagation signals; and multimodal augmentation enables reasoning over text, images, videos, and cross-modal inconsistencies. 
\item[-] Judgment-oriented studies place LLMs closer to the final decision process. 
Instruction prompting reformulates misinformation detection as reliability assessment, while supervised and reinforcement tuning adapt detectors to specific domains, evolving events, adversarial samples, or stance-aware debunking. 
Explanation-aware judgment further extends outputs from labels to rationales, multimodal explanations, and corrective responses. 
\item[-] For RAG-based systems, knowledge grounding and evidence retrieval reduce reliance on parametric knowledge by introducing external and dynamically updated information. 
Evidence reasoning then determines whether retrieved evidence truly supports or refutes a claim. 
Agentic methods formulates fact-checking and detection system into well-designed multi-step workflows such as claim decomposition, evidence search, and debunking. 
\end{itemize}
Overall, LLM-based defense broadens misinformation detection into a more comprehensive verification pipeline that combines representation enrichment, natural-language judgment, evidence grounding, and agentic reasoning.

\section{Countermeasures: Adversarial Robustness against LLM-enabled Misinformation Threats} \label{sec:countermeasures}
LLM-enabled misinformation attacks expand from content generation to social-context manipulation and evidence poisoning.
In this section, we investigate existing countermeasures at different layers of the misinformation pipeline.
This view highlights both the progress and the uneven coverage of current defenses.

\begin{table*}[t]
\centering
\small
\caption{Attack vs. defense correspondence matrix for LLM-enabled misinformation threats at content level.}
\label{tab:attack_defense_matrix_content}
\resizebox{\linewidth}{!}{
\begin{tabular}{p{2.1cm} p{6.5cm} p{6cm}}
\toprule
  \multicolumn{1}{c}{\textbf{Attack Type}} & \multicolumn{1}{c}{\textbf{Representative Methods on Attacks}} & \multicolumn{1}{c}{\textbf{Representative Methods on Defenses}}  \\
\midrule
Synthetic News Generation &
\makecell[lt]{Prompt-based fake news generation \\ \cite{pan2023risk,lucas2023fighting,su2023fake,sallami2024deception,wang2025have,vykopal2024disinformation,chen2024can}; \\ 
SFT-based domain-adapted generation \cite{puccetti2024ai}} 
& \makecell[lt]{LLM-generated Text Detection        \cite{wang2026prompt,nathanson2024step,su2024adapting,beigi2024model}; \\ 
Semantic/Symbolic adversarial learning  \cite{ding2025dynamic,tian2025symbolic}; \\
Multi-agent mitigation \cite{aldahoul2025toward}} \\
\cmidrule{2-3}  
& 
Retrieval-grounded generation \cite{singh2024adversarial} 
& $\times$
\\ 
\midrule
{Persuasive Targeting} &
\makecell[lt]{Personalization \cite{zugecova2025evaluation,zhao2025lance}; \\ Emotional prompting \cite{vinay2025emotional}} &
$\times$ \\
\midrule 
Content Camouflage and Evasion &
\makecell[lt]{Detector evasion via paraphrasing \cite{das2025fake,sakib2026credibility}, \\  style transfering \cite{wu2024fake},  sentiment \\ manipulation \cite{tahmasebi2026robust},  prompt optimization \cite{lularge2024}}& 
\makecell[lt]{Style-attack adversarial learning \cite{park2025adversarial,wu2024fake}; \\  
 Content-style Disentanglement \cite{vu2026enhancing}; \\ 
 Invariant Representation Learning    \cite{fei2026enhancing}   
 } \\ 
\cmidrule{2-3}  
& \makecell[lt]{Adversarial attack via prompting \cite{deverna2024fact,danry2025deceptive,xu2025ssa,przybyla2025attacking}, \\  entropy-based \cite{lu2026llm} and RL-based perturbations \cite{chen2023anti}
}
& \makecell[lt]{Adversarial training \cite{aldahoul2025toward,ding2025dynamic,tian2025symbolic};\\
Invariant Representation Learning    \cite{fei2026enhancing}    
} \\
\cmidrule{2-3}
&  Jailbreak attack \cite{kaneko2026jailnewsbench} & $\times$  \\ 
\midrule
Scenario-specific Ecosystem & 
\makecell[lt]{News recommendation system:\\ training data pollution
\cite{hu2025llm}, \\ and RL-based ranking manipulation \cite{zhao2025lance}}& $\times$  \\
\cmidrule{2-3}  
& \makecell[lt]{Knowledge-intensive task: evidence \\ pollution \cite{pan2023risk,pan2023attacking}} & $\times$  \\
\bottomrule
\multicolumn{3}{l}{\footnotesize{``$\times$'' indicates that we did not identify directly targeted defenses in the misinformation-specific literature.}} 
\end{tabular}
}
\end{table*}

\subsection{Robustness against Content-level Attacks}
\label{sec:content_defense}
Content-level defenses address attacks that manipulate the misinformation artifact itself, including synthetic news generation, persuasive targeting, content camouflage and evasion, and scenario-specific ecosystem attacks (see Sec. \ref{sec:attack_content} for details). 
Existing studies mainly explore LLM-generated text detection, adversarial training, invariant representation learning, and multi-agent mitigation.

\subsubsection{LLM-generated text detection}
LLM-generated text detection provides a direct defense against synthetic misinformation by identifying whether a news article, claim, or post has been generated or rewritten by LLMs. 
Existing methods model prompt-induced linguistic fingerprints, generation traces, or distributional differences between human-written and LLM-generated misinformation \cite{wang2026prompt,nathanson2024step}. 
Recent work studies detection under LLM-induced distribution shifts, including mixtures of human-written, LLM-generated, and LLM-paraphrased content~\cite{su2024adapting}, as well as source-model attribution~\cite{beigi2024model}. 
These defenses provide a natural first step for detecting synthetic misinformation, although their effectiveness depends on whether generation artifacts remain detectable.

\subsubsection{Adversarial training}
Adversarial training methods improve detector robustness by exposing models to LLM-generated or LLM-rewritten attack variants during training. 
For synthetic news generation threat, semantic and symbolic adversarial learning train detectors on challenging generated variants and semantically constrained fake information \cite{ding2025dynamic,tian2025symbolic}. 
For content camouflage and evasion attack, some works expose detectors to style-transferred or style-camouflaged misinformation, encouraging them to rely less on tone, fluency, or news-like presentation \cite{park2025adversarial,wu2024fake}.

\subsubsection{Content-style disentanglement and invariant representation learning}
Content-style disentanglement and invariant representation learning address scenarios in which misinformation is rewritten into different surface forms while preserving its misleading semantics. 
\citet{vu2026enhancing} separate factual or deceptive semantics from writing style, sentiment, credibility cues, and presentation. 
\citet{fei2026enhancing} further study invariant representation learning that aligns original and adversarially transformed versions, enabling detectors to focus on stable claim-level signals. 

\subsubsection{Multi-agent mitigation}
Multi-agent mitigation methods use multiple LLM agents or roles to cross-check suspicious content, decompose claims, challenge weak reasoning, and aggregate different perspectives. 
\citet{aldahoul2025toward} design multi-agent systems to combine claim understanding, evidence seeking, stance analysis, and final judgment for misinformation mitigation. 
Multi-agent designs can improve coverage and reasoning diversity, but they also introduce coordination costs, disagreement-resolution problems, and the risk of error propagation across agents.

\paragraph{Discussion: Gap with respect to content-level attacks}
As shown in Table~\ref{tab:attack_defense_matrix_content}, existing content-level defenses are concentrated on synthetic misinformation generation and content camouflage or evasion. 
A key remaining challenge is to bridge textual robustness and downstream application robustness, especially when manipulated content is personalized, retrieved, recommended, or reused by knowledge-intensive systems.

\begin{table*}[t]
\centering
\small
\caption{Attack vs. defense correspondence matrix for LLM-enabled misinformation threats at social level.}
\label{tab:attack_defense_matrix_social}
\resizebox{\linewidth}{!}{
\begin{tabular}{p{2.5cm} p{6.2cm} p{5.5cm}}
\toprule
  \multicolumn{1}{c}{\textbf{Attack Type}} & \multicolumn{1}{c}{\textbf{Representative Methods on Attacks}} & \multicolumn{1}{c}{\textbf{Representative Methods on Defenses}}  \\
\midrule
Comment Manipulation &
\makecell[lt]{
Narrative-level amplification via prompt-based \\ methods \cite{bandara2024hallucination,kim2025breaking}}  & CoT Prompting \cite{wu2025chainofethics} \\ 
\cmidrule{2-3}
& \makecell[lt]{Malicious Comment generation via prompt-based \\  methods \cite{huynh2024improving,ahmed2026new} and self-reflection \cite{underwood2026generating}; \\ 
Message injection attack \cite{luo2024message}} &   
Invariant representation learning \cite{zhang2025sincon}, and manipulation identification \cite{shen2025llm}  \\
\cmidrule{2-3}
&  Multi-round fake narrative generation \cite{brian2025mpcg,liu2025stepwise} & $\times$ \\ 
\cmidrule{2-3}
& \makecell[lt]{Real-comment retrieval attack \cite{koren2025evaluating}} & $\times$   \\ 
\midrule
Structure Manipulation & 
\makecell[lt]{User engagement manipulation \cite{wang2023attacking,wang2024bots}, \\ and fake interaction attack \cite{zhu2024general}} & Structure uncertainty modeling \cite{zeng2025robustness} \\  
\midrule
Strategic Campaign Manipulation & \makecell[lt]{RL-based comment/bot control \cite{le2022socialbots,yang2025robctrl,tong2025group}; \\ 
Campaign simulation \cite{qiao2025botsim,qiao-etal-2025-dynamic,lu2026large}; \\ 
Actor identity manipulation \cite{feng2024does}} & $\times$  \\
\midrule
Propagation Data Poisoning & \makecell[lt]{Backdoor attack \cite{jin2025backdoor,goschprovable}; \\ 
Platform data poisoning \cite{yamashita2024fake,wu2024attacking}}  & $\times$
\\ 
\bottomrule
\multicolumn{3}{l}{\footnotesize{``$\times$'' indicates that we did not identify directly targeted defenses in the misinformation-specific literature.}} 
\end{tabular}
}
\end{table*}

\subsection{Robustness against Social-level Attacks}
\label{sec:social_defense}

Social-level defenses address attacks that manipulate the social context surrounding misinformation as shown in Sec.~\ref{sec:attack_social}. 
Existing studies can be mainly divided into prompting-based reasoning, invariant representation learning, manipulation identification, and structure uncertainty modeling.

\subsubsection{Prompting-based reasoning}
Prompting-based defenses use LLM reasoning to identify manipulative, unethical, or coordinated narratives in social discussions. 
Chain-of-thought and ethics-oriented prompting can help detect problematic narratives in LLM-generated comments, replies, or discussion threads \cite{wu2025chainofethics}. 
These methods mainly correspond to narrative-level manipulation, where misinformation is strengthened through surrounding discussion rather than by changing the original article. 
They provide a flexible reasoning-based defense, but their reliability depends on prompt design and may vary across personas, tones, and conversational strategies.

\subsubsection{Invariant representation learning for manipulated social context}
Invariant representation learning aims to reduce detector sensitivity to manipulated comments, messages, or propagation contexts. 
Existing studies learn stable representations under LLM-generated narratives or adversarially changed social signals \cite{zhang2025sincon}.

\subsubsection{Manipulation identification}
Manipulation identification methods explicitly detect whether the social context itself has been adversarially injected or strategically manipulated. 
Rather than directly predicting the veracity of a news item, \citet{shen2025llm} examine whether comments, replies, or discussion patterns show abnormal or manipulated characteristics.
This direction is particularly relevant to message injection and malicious comment generation, where the attacker targets the detector's contextual input rather than the misinformation content. 

\subsubsection{Structure uncertainty modeling}
Structure uncertainty modeling addresses attacks on propagation graphs, user-news interactions, and engagement patterns. 
\citet{zeng2025robustness} model user-news interactive edges as noisy and potentially adversarial rather than fully reliable. 
They can alleviate local graph or engagement manipulation but usually fail under a long-horizon campaign behavior.

\paragraph{Discussion: Gap with respect to social-level attacks}
As shown in Table~\ref{tab:attack_defense_matrix_social}, current social-level defenses mainly cover narrative-level amplification, malicious comment generation, message injection, and structure manipulation. 
Direct countermeasures remain limited for multi-round fake narrative generation, real-comment retrieval attacks, strategic campaign manipulation, and propagation data poisoning. 
These attacks differ from localized comment or graph perturbations because they may involve temporal adaptation, coordinated actors, real user comments, bot control, identity manipulation, or contaminated platform data.

\begin{table*}[t]
\centering
\small
\caption{Attack vs. defense correspondence matrix for LLM-enabled misinformation threats at evidence level.}
\label{tab:attack_defense_matrix_evidence}
\resizebox{\linewidth}{!}{
\begin{tabular}{p{2cm} p{6.4cm} p{5.6cm}}
\toprule
  \multicolumn{1}{c}{\textbf{Attack Type}} & \multicolumn{1}{c}{\textbf{Representative Methods on Attacks}} & \multicolumn{1}{c}{\textbf{Representative Methods on Defenses}}  \\
\midrule 
Evidence Content Manipulation  &
Synthetic or fabricated evidence generation attacks via prompt-based methods \cite{abdelnabi2023fact,du2022synthetic,yang2026steering}
& 
Evidence reliability modeling
\cite{turaga2024information,wan2026facade}; and structured evidence reasoning and causal/multi-hop verification
\cite{habib2026llm,zhang2024causal}    \\
\cmidrule{2-3}
& Adversarial claim attacks via prompt optimization \cite{bethany2025camouflage}, iterative transformation \cite{islam2025inconsistent}, fine-tuning \cite{layne2025analyzing}, persuasion \cite{leite2026llm}, and search-enabled agentic pipeline \cite{ou2026deceive} & Evidence-grounded verification \cite{xiang2024certifiably,ma2026radar}, and multi-agent LLMs \cite{aldahoul2025toward}
 \\
 \midrule
Evidence Corpus Poisoning &  \makecell[lt]{Targeted poisoning attack via heuristic prompting \\ methods \cite{zou2025poisonedrag,li2025cpa,song2025chain}, agentic claim decomposition  \\ simulation \cite{he2026fact2fiction}, 
gradient-aware adversarial \\ optimization \cite{gong2025topic}, and semantic alignment  \cite{wu2025admit}}
& \makecell[lt]{Evidence trustworthy reasoning \cite{habib2026llm,zhang2024causal}, and  \\ poisoning attribution via prompting  \cite{zhang2025traceback,zhang2025taught} \\ and LLM activation analysis  \cite{tan-etal-2025-revprag}} \\
\cmidrule{2-3}
& More efficient poisoning attack (e.g., single-document poisoning) via chain-of-evidence \cite{chang-etal-2025-one,zhang2026practical,song2026adversarialcot}& $\times$ \\ 
\cmidrule{2-3}
& Universal attacks for cross-topic queries \cite{geng2025unic}, query-agnostic \cite{chen2025mirage}& $\times$ \\
 \cmidrule{2-3}
 & Multiple competing poisoning attacks \cite{chen2025poisonarena} & $\times$ \\
 \cmidrule{2-3}
&  Structured knowledge poisoning and Trojan-style RAG attacks
\cite{chen2026kepo,cheng2024trojanrag} &  $\times$  \\
 \midrule
Evidence Retrieval Manipulation &
\makecell[lt]{Retrieval visibility manipulation 
\cite{cho2024typos,song2025silent,chang2026overcoming}, \\ blocker documents and retrieval jamming that \\ suppress relevant evidence
\cite{shafran2025machine}, and surrogate \\ retriever via reverse engineering \cite{chen2025flippedrag}} & $\times$  \\ 
\bottomrule
\multicolumn{3}{l}{\footnotesize{``$\times$'' indicates that we did not identify directly targeted defenses in the misinformation-specific literature.}} 
\end{tabular}
}
\end{table*}

\subsection{Robustness against Evidence-level Attacks}
\label{sec:evidence_defense}
Evidence-level defenses protect fact-checking systems, RAG pipelines, and search-augmented LLM agents from attacks on evidence content, evidence corpora, and evidence retrieval. 
Existing studies mainly examine evidence reliability modeling, structured or causal evidence reasoning, evidence-grounded verification, multi-agent verification, and poisoning attribution.

\subsubsection{Evidence reliability modeling}
Evidence reliability modeling evaluates whether retrieved evidence is trustworthy rather than merely relevant. 
Existing studies model evidence reliability for misinformation detection and investigate credibility signals that help assess external sources and retrieved content \cite{turaga2024information}. 
Other work studies how deceptive evidence influences LLM-based fact-checking and proposes mitigation strategies against misleading evidence use \cite{wan2026facade}. 

\subsubsection{Structured evidence reasoning and causal or multi-hop verification}
Structured evidence reasoning examines whether evidence supports, contradicts, or contextualizes a claim. 
It shifts evidence defense from retrieval alone to verification of the relation between claim and evidence.
Existing studies use LLM-powered reasoning to integrate multimodal evidence for fake-news detection \cite{habib2026llm} and causal or multi-hop verification to reduce reliance on spurious evidence associations \cite{zhang2024causal}. 
These methods require the system to assess claim-evidence relations rather than accept surface-level semantic overlap.

\subsubsection{Evidence-grounded verification and multi-agent LLMs}
Evidence-grounded verification explicitly links final judgments to retrieved, checked, or refined evidence. 
\citet{xiang2024certifiably} use an isolate-then-aggregate strategy to reduce the influence of poisoned evidence on the final answer.
\citet{ma2026radar} improve robustness against generated evasion samples via retrieval-augmented fake-news detection with adversarial refinement. 
\citet{aldahoul2025toward} explore multi-agent LLM systems to decompose verification into claim analysis, evidence retrieval, evidence critique, and final judgment for robust detection. 
These methods provide a bridge between evidence content defenses and workflow-based verification, although they inherit dependencies on retrieval quality, source reliability, and tool use.

\subsubsection{Poisoning attribution and secure RAG analysis}
Poisoning attribution addresses evidence corpus poisoning by tracing corrupted RAG outputs back to suspicious documents or knowledge sources. 
Prompting-based attribution methods identify retrieved or stored documents that may be responsible for poisoned generations \cite{zhang2025traceback,zhang2025taught}. 
Activation-based analysis further detects poisoned RAG responses when malicious evidence is not apparent from surface text \cite{tan-etal-2025-revprag}. 
These methods mainly provide diagnostic and attribution tools for poisoned retrieval settings.

\paragraph{Discussion: Gap with respect to evidence-level attacks}
As shown in Table~\ref{tab:attack_defense_matrix_evidence}, current evidence-level defenses primarily counter fabricated evidence, adversarial claim attacks, and targeted corpus poisoning. 
These defenses remain insufficient against evidence-layer manipulation, including efficient single-document or chain-of-evidence poisoning, query-agnostic universal poisoning, competing poisoning, structured knowledge poisoning, and evidence retrieval manipulation.
These settings move beyond the attack on a known target claim toward cross-topic, query-agnostic, and competing evidence manipulation.
Existing reliability modeling and attribution methods may provide partial diagnostic signals, but direct countermeasures are not yet systematically established.

\subsection{Summary}
\label{sec:defense_summary}
Overall, existing countermeasures reveal a clear mismatch between the maturity of defenses and the frontier of LLM-enabled misinformation attacks. 
Current defenses mainly cover localized and pre-defined attacks, whereas emerging threats are becoming increasingly adaptive, knowledge-aware, long-horizon, and stealthy. 
This gap highlights the need for more systematic defenses against emerging attack paradigms, particularly coordinated misinformation campaigns and covert attacks targeting evidence-grounded systems. 

\section{Datasets and Evaluation Metrics}

\begin{table}[t]
\centering
\small
\caption{Summary of representative misinformation datasets. MM, Cmt, Rel, Tem, GE refer to multimodal, comment, relations, temporal information, and gold evidence. }
\label{tab:dataset_taxonomy}
\resizebox{\linewidth}{!}{
\begin{tabular}{l|c|cc|ccc|cl}
\toprule
\multicolumn{1}{c|}{\multirow{2}{*}{Dataset}} & \multicolumn{1}{c|}{\multirow{2}{*}{\makecell{Target\\Source}}} & \multicolumn{2}{c|}{Content-centric}  & \multicolumn{3}{c|}{Social-context}  & \multicolumn{2}{c}{Evidence-grounded}  \\ 
&   & Text & MM & Cmt & Rel & Tem & GE  &  Evidence Source   \\ 
\midrule
Mocheg \cite{yao2023end} & Human  
& \checkmark & \checkmark & -- & -- & -- & -- & -- \\
NewsCLIPpings \cite{luo2021newsclippings} & Human & \checkmark & \checkmark & -- & -- & -- & -- & --   \\ 
Fakeddit \cite{nakamura2020fakeddit} & Human & \checkmark & \checkmark & -- & -- & -- & -- & -- \\ 
Twitter15/Twitter16 \cite{ma2017detect} & Human
& \checkmark & -- & \checkmark & \checkmark & \checkmark 
& -- & -- \\
Weibo \cite{ma2016detecting} & Human
& \checkmark & \checkmark & \checkmark & \checkmark & \checkmark
& -- & -- \\
Weibo21 \cite{nan2021mdfend} & Human
& \checkmark & \checkmark & \checkmark & -- & --
& -- & -- \\ 
PHEME \cite{zubiaga2016learning,kochkina2018pheme} & Human
& \checkmark & -- & \checkmark & \checkmark & \checkmark 
& -- & -- \\
PHEMEPlus \cite{dougrez2022phemeplus} & Human
& \checkmark & -- & \checkmark & \checkmark & \checkmark 
& $\checkmark$ & Web Search \\ 
FakeNewsNet \cite{shu2020fakenewsnet} & Human
& \checkmark & -- & \checkmark & \checkmark & \checkmark
& -- & -- \\
FakeHealth \cite{dai2020ginger} & Human & \checkmark & \checkmark & \checkmark & \checkmark & \checkmark
& -- & -- \\ 
CoAID \cite{DBLP:journals/corr/abs-2006-00885} & Human
& \checkmark & -- & \checkmark & \checkmark & \checkmark
& -- & -- \\ 

LIAR \cite{wang2017liar} & Human
& \checkmark & -- & -- & -- & -- 
& -- & -- \\
BUZZFEEDNEWS \cite{potthast2018stylometric} & Human
& \checkmark & -- & -- & -- & -- & -- & --  \\ 
BUZZFACE \cite{DBLP:conf/icwsm/SantiaW18} & Human
& \checkmark & -- & \checkmark & -- & -- & -- & -- \\  
FEVER \cite{thorne2018fever} & Human
& \checkmark & -- & -- & -- & -- 
& \checkmark & Wikipedia \\

WICE \cite{kamoi2023wice} & Human
& \checkmark & -- & -- & -- & -- 
& \checkmark & Wikipedia \\

FEVEROUS \cite{aly2021fact} & Human
& \checkmark & -- & -- & -- & -- 
& \checkmark & Wikipedia \\

SciFact \cite{wadden2020fact} & Human
& \checkmark & -- & -- & -- & -- 
& \checkmark & Scientific Corpus \\

\midrule 
LLMFake \cite{chen2024can} & 7 LLMs 
& \checkmark & --   & -- & --  & -- 
& -- & -- \\ 
Grover \cite{DBLP:conf/nips/ZellersHRBFRC19} & GPT-2  
& \checkmark & -- & -- & --  & -- 
& -- & -- \\ 
UHGEval \cite{DBLP:conf/acl/LiangSNLXTWHPWD24} & 5 LLMs  & \checkmark & -- & -- & --  & -- 
& -- & --  \\ 

FacTool \cite{chern25factool} & ChatGPT
& \checkmark & -- & -- & -- & -- 
& --  & --  \\

HaluEval \cite{li2023halueval} & ChatGPT
& \checkmark & -- & -- & -- & --  
& -- & -- \\

FELM \cite{zhao2023felm} & ChatGPT
& \checkmark & -- & -- & -- & -- 
& \checkmark & Google Search \\

FActScore \cite{min2023factscore} & 3 LLMs 
& \checkmark & -- & -- & -- & -- 
& \checkmark & Wikipedia \\

FactCheck-GPT \cite{wang2024factcheck} & 2 LLMs
& \checkmark & -- & -- & -- & -- 
& \checkmark & Google Search \\

BingCheck \cite{li2024self} & Bing Chats
& \checkmark & -- & -- & -- & -- 
& \checkmark & Bing Search \\
MFC-Bench \cite{wang2025mfc} & 5 LVLMs & \checkmark &  \checkmark & -- & -- & --  & -- & --\\ 
DGM4 \cite{shao2023detecting} & SLMs 
& \checkmark &  \checkmark & -- & -- & --  & -- & -- \\
LiveFact \cite{xu2026livefact} & Qwen3-235B-A22B & \checkmark & -- & -- & -- & -- & \checkmark & Google Search \\ 
\midrule
NQ \cite{kwiatkowski2019natural} & --
& \checkmark & -- & -- & -- & -- 
& -- & Wikipedia \\

HotpotQA \cite{yang2018hotpotqa} & --
& \checkmark & -- & -- & -- & -- 
& \checkmark & Wikipedia \\

MS-MARCO \cite{nguyen2016ms} & --
& \checkmark & -- & -- & -- & -- 
& -- & Web \\

SQuAD \cite{rajpurkar2016squad} & --
& \checkmark & -- & -- & -- & -- 
& \checkmark & Wikipedia \\

BoolQ \cite{clark2019boolq} & --
& \checkmark & -- & -- & -- & -- 
& \checkmark & Wikipedia \\

BBQ \cite{parrish2022bbq} & --
& \checkmark & -- & -- & -- & -- 
& -- & -- \\

StereoSet \cite{nadeem2021stereoset} & --
& \checkmark & -- & -- & -- & -- 
& -- & -- \\

GraphRAG-Bench \cite{xiang2025use} & --
& \checkmark & -- & -- & \checkmark & -- 
& \checkmark & Documents \\

MuSiQue \cite{trivedi2022musique} & --
& \checkmark & -- & -- & -- & -- 
& \checkmark & Wikipedia \\
\bottomrule  
\end{tabular}
}
\end{table}

This section summarizes representative misinformation datasets and metrics.

\subsection{Datasets} 

Existing datasets differ mainly in what information they provide for evaluation: content, social context, and evidence. Table~\ref{tab:dataset_taxonomy} follows this distinction and lists representative datasets rather than an exhaustive catalog.
Traditional datasets mainly support content- and social-context-aware detection, while evidence-grounded datasets support claim verification and fact-checking. 
LLM-era datasets focus on machine-generated misinformation, and factuality evaluation. 
Most current resources still evaluate only one or two signals at a time.

\subsubsection{Human-generated Datasets}
Human-generated datasets are constructed from political claims, news articles, social media posts, user comments, propagation traces, or manually curated evidence. 
Content-centric datasets evaluate whether a model can identify false or misleading information from textual or multimodal content. 
Representative datasets include Weibo21, BuzzFeedNews~\cite{potthast2018stylometric}, BuzzFace~\cite{DBLP:conf/icwsm/SantiaW18}, Fakeddit~\cite{nakamura2020fakeddit}, and NewsCLIPpings~\cite{luo2021newsclippings}. 
These datasets range from article-level fake-news classification to multimodal and out-of-context image--caption verification, reflecting the shift from text-only misinformation to multimodal manipulation.
Social-context datasets incorporate user reactions, comments, reposts, temporal cascades, or propagation structures. 
Representative datasets include Twitter15 and Twitter16~\cite{ma2017detect}, Weibo~\cite{ma2016detecting}, PHEME~\cite{zubiaga2016learning,kochkina2018pheme}, PHEMEPlus~\cite{dougrez2022phemeplus}, FakeNewsNet~\cite{shu2020fakenewsnet}, FakeHealth~\cite{dai2020ginger}, and CoAID~\cite{DBLP:journals/corr/abs-2006-00885}. 
Evidence-grounded datasets require systems to verify claims using external evidence rather than relying only on content features. 
Representative datasets include LIAR~\cite{wang2017liar}, FEVER~\cite{thorne2018fever}, WICE~\cite{kamoi2023wice}, FEVEROUS~\cite{aly2021fact}, SciFact, and Mocheg~\cite{yao2023end}. 
These datasets support evidence retrieval, rationale selection, multimodal evidence reasoning, and temporally aware verification.

\subsubsection{LLM-era Datasets}
This category contains machine-generated misinformation. 
Representative datasets include LLMFake~\cite{chen2024can}, Grover~\cite{DBLP:conf/nips/ZellersHRBFRC19}, MFC-Bench~\cite{wang2025mfc}, and DGM4~\cite{shao2023detecting}. 
They evaluate whether detectors can handle LLM-generated false articles, controllable fake news, multimodal fact-checking samples, and generative image--text manipulations. 
Compared with human-generated datasets, this category is still relatively limited, revealing a gap in realistic LLM-era misinformation benchmarks.
QA and RAG benchmarks such as NQ~\cite{kwiatkowski2019natural}, HotpotQA~\cite{yang2018hotpotqa}, MS-MARCO~\cite{nguyen2016ms}, SQuAD~\cite{rajpurkar2016squad}, BoolQ~\cite{clark2019boolq}, MuSiQue~\cite{trivedi2022musique}, GraphRAG-Bench~\cite{xiang2025use}, BBQ~\cite{parrish2022bbq}, and StereoSet~\cite{nadeem2021stereoset} can also serve as diagnostic benchmarks for retrieval quality, evidence reasoning, and bias under uncertainty.
Some datasets evaluate factuality in LLM outputs including HaluEval~\cite{li2023halueval}, FELM~\cite{zhao2023felm}, FacTool~\cite{chern25factool}, FActScore~\cite{min2023factscore}, FactCheck-GPT~\cite{wang2024factcheck}, and BingCheck~\cite{li2024self}. 

\subsection{Evaluation Metrics}
Metrics in LLM-era misinformation evaluation measure not only clean detection performance, but also whether the system retrieves trustworthy evidence, resists adversarial manipulation, and preserves utility after defense. 
We summarize representative metrics in Table~\ref{tab:metric_taxonomy}. 

\begin{table*}[t]
    \centering
    \small
    \caption{Metrics for evaluating LLM-era misinformation systems.}
    \label{tab:metric_taxonomy}
  \resizebox{\linewidth}{!}{
  \begin{tabular}{lp{5cm}p{6.9cm}}
    \toprule
    \textbf{Evaluation goal} & \textbf{What it measures} & \textbf{Representative metrics} \\
    \midrule
    Task performance &
    Whether the system predicts correct labels under benign conditions &
    Accuracy, precision, recall, F1 of positive class, macro-F1, false positive rate, false negative rate, and AUC \\
    \midrule
    Evidence quality &
    Whether the verdict is supported by relevant and sufficient evidence rather than unsupported rationales &
    Evidence recall/precision, FEVER \cite{thorne2018fever}, Oracle FEVER \cite{nie2019combining}, claim metrics \cite{ullrich2025claim}, evidence sufficiency \cite{atanasova2022fact}, rationale faithfulness \cite{deyoung2020eraser}, factual consistency \cite{kryscinski2020evaluating}, and atomic factual precision \cite{min2023factscore}  \\
    \midrule
    Attack effectiveness &
    Whether an adversary can flip labels, inject poisoned evidence, manipulate retrieval, or produce deceptive rationales &
    Attack success rate (ASR) \cite{zou2025poisonedrag,luo2024message,aldahoul2025toward}, multi-attack ASR \cite{chen2025poisonarena}, top-k adversarial document retrieval rate \cite{wang2025bias}, poisoned evidence recall/precision \cite{ha2025mm}, ranking boost  \cite{ha2025mm}, deceived justification rate \cite{wu2025admit}, perplexity \\
    \midrule
    Defense robustness &
    Whether a mitigation lowers attack success while preserving benign utility &
    Accuracy/F1 under attack, and utility-robustness trade-off \cite{zhang2019theoretically}  \\
    \midrule
    Deployment impact &
    Whether the system is practical and usable in fact-checking workflows &
    Runtime \cite{zou2025poisonedrag}, number of queries \cite{zou2025poisonedrag} \\
    \bottomrule
    \end{tabular}
    }
\end{table*}

\subsubsection{Task Performance Metrics}
Task performance metrics measure whether a system can predict correct labels under benign conditions, such as accuracy, precision, recall, macro-F1, false positive rate, false negative rate and AUC etc. 

\subsubsection{Evidence and Factuality Metrics}
Evidence-aware metrics evaluate whether a system retrieves and uses the information required to justify its verdict. Evidence recall and precision measure whether relevant sources are successfully retrieved, whereas FEVER-style scores require both a correct label and sufficient supporting evidence. Evidence quality also include sufficiency, rationale faithfulness, factual consistency, and atomic factual precision, because a fluent explanation may appear persuasive even when it is not grounded in the retrieved evidence.

\subsubsection{Attack Effectiveness Metrics}
Attack metrics quantify the extent to which an adversary achieves the intended manipulation. Output-level metrics such as attack success rate and multi-attack success rate capture whether the final prediction or rationale has been altered. Retrieval-level metrics such as adversarial retrieval rate, poisoned evidence recall, and ranking boost should be reported separately, because a poisoned document may be retrieved without affecting the final verdict, or may change the verdict only through deceptive justification.

\subsubsection{Defense Robustness Metrics}
Defense evaluation should report both utility and robustness. Accuracy under attack and robustness gap summarize how much performance degrades under adversarial inputs or poisoned evidence. Defense success rate and utility--robustness trade-off further indicate whether a mitigation blocks attacks without sacrificing too much clean performance. This paired reporting prevents defenses from appearing effective merely because they reject many inputs, over-filter evidence, or reduce the usefulness of the underlying fact-checking system.

\section{Challenges and Future Directions}
We discuss the key challenges around evaluation, system robustness, and deployment.

\subsection{Evaluation Gap: From Detection Accuracy to Budgeted Risk Evaluation}
The evaluation gap concerns how to measure the LLM-enabled misinformation risk under realistic adversarial conditions. This gap calls for a shift from isolated detection accuracy to budgeted ecosystem-level risk assessment.

\begin{itemize}
\item[-] \textbf{Multi-dimensional risk measurement.}
Most existing studies use attack success rate as the primary metric. 
However, the risks posed by LLM-enabled misinformation are not limited to detector evasion. 
An adversarial example may preserve the misleading claim, increase its visibility, influence user beliefs, or distort retrieved evidence. 
Broader risk indicators are needed to cover more diverse evaluation dimensions.

\item[-] \textbf{Practical threat models with limited attacker access and budget.}
Many attacks are evaluated under relatively strong assumptions such as white-box access, sufficient query budgets, or repeated feedback from the target system.
In practice, misinformation actors often operate with partial knowledge, limited feedback, and constrained resources.
Thus, black-box, low-budget, and feedback-limited settings are important for assessing whether robustness holds under practical attack constraints.

 \item[-] \textbf{Complex multilingual and multimodal attack evaluation.}
False or misleading narratives may combine generated claims with reused images, synthetic visuals, manipulated captions, or short videos. 
More comprehensive evaluation benchmarks are needed to assess whether detection and verification systems remain reliable when misleading cues are distributed across textual claims, image-text relationships, and video evidence.

\item[-] \textbf{Long-horizon propagation risk evaluation.}
Current works have shown that LLMs can manipulate comments, personas, and stance signals during propagation. 
As recent agentic systems become capable of simulating social interaction and propagation dynamics, which may be misused to generate more adaptive campaigns, long-horizon evaluation becomes a key setting for studying self-evolving misinformation campaigns.

\item[-] \textbf{Competing and multi-objective attack evaluation.}
Many existing attacks optimize a single objective such as evading a detector or increasing the ranking of poisoned evidence. 
Realistic attacks, however, often involve multiple objectives, which can conflict: aggressive rewriting may improve evasion but reduce credibility, whereas coordinated amplification may increase exposure but also make the campaign easier to flag. 
This makes multi-objective evaluation an urgent need for characterizing competing or collaborative attack behaviors.

\end{itemize}

\subsection{System Gap: Hardening LLM-based Verification and Information Systems}

The system gap concerns how to make LLM-based misinformation defenses robust when the system itself becomes part of the attack surface. 

\begin{itemize}
\item[-] \textbf{Robustness of LLM-based detectors and agentic workflow.}
Research on LLM-based detectors  and agentic workflow creates attack surfaces that are absent or less explicit in conventional classifiers, such as prompt injection, tool-use manipulation, and memory poisoning. 
Future work needs therefore analyze to the security of the full verification workflow and build more robust LLM-centered and agentic systems. 

\item[-] \textbf{Content-level defense against personalized misinformation attacks.}
Unlike earlier forms of misinformation generation, LLM-enabled attacks can adapt claims to specific audiences, rewrite them to evade generic classifiers, and use rhetorical strategies that increase perceived credibility.
This makes it harder to detect with a generic classifier.
Future defenses need robustness against adaptive paraphrasing and persuasion-preserving transformations. 

\item[-] \textbf{Defense against evolving LLM-enabled social manipulation attacks.}
LLMs enable social-level adversaries to coordinate engagement and manipulate propagation patterns in a strategic manner.  Unlike traditional heuristic perturbations, these attacks can dynamically adapt to evade propagation-based detectors. A key future direction is to develop defenses against stealthier and more adaptive LLM-driven propagation manipulation.

\item[-] \textbf{Evidence integrity and provenance modeling for robust RAG.}
Existing RAG-based defenses often emphasize retrieval relevance, but relevance alone does not guarantee source authenticity. 
Robust evidence integrity and provenance modeling are therefore important for distinguishing reliable grounding from adversarially constructed support.

\item[-] \textbf{Unintentional misinformation from personalized LLM systems.}
Not all LLM-related misinformation comes from deliberate attacks. Personalized assistants may generate misleading information when they rely on outdated memory or over-personalized assumptions. Such errors may arise from both hallucinations and omissions of relevant facts. 
The factual reliability of personalized LLM systems is therefore an important part of misinformation defense, especially for time-sensitive or high-impact information needs.

\end{itemize}

\subsection{Deployment Gap: Operationalizing Robust, Governed, and Human-centered Defense}
The deployment gap concerns how LLM-based misinformation defenses can be used in real environments with limited resources,  privacy constraints, platform-specific norms, and human oversight.

\begin{itemize}
\item[-] \textbf{Risk-adaptive defense under practical resource constraints.}
Many LLM-based defenses rely on computationally expensive verification mechanisms, which are often impractical for online deployment. 
Future systems should allocate verification resources in a tiered manner, so that high-risk cases receive stronger evidence verification, provenance validation, and human oversight.
\item[-] \textbf{Governed use of social signals and human oversight.}
Social-level defenses often rely on account behavior, engagement traces, and coordination patterns. 
These signals help detect manipulation and misinformation but also raise concerns regarding privacy and fairness. 
Future deployment should distinguish manipulative coordination from legitimate collective behavior and integrate human review for ambiguous or high-impact cases.
\item[-] \textbf{Auditable decision-making and post-deployment monitoring.}
LLM-based defenses may label, downrank, or ignore content based on model judgments and platform rules. 
Without auditability, it is difficult to diagnose false positives, false negatives, delayed interventions, or decisions affected by poisoned evidence. 
Future systems should record key decision traces and monitor long-term drift, attacker adaptation, and downstream effect.
\end{itemize}

\section{Conclusion}
This paper presents a tri-role perspective on the evolving relationship between large language models and misinformation. Rather than viewing LLMs solely as generators of synthetic false content, we argue that they should be understood as attackers, victims, and defenders within a broader misinformation ecosystem. We systematically analyze how LLMs empower misinformation attacks across content, social, and evidence layers; how LLM-based detection paradigms themselves become vulnerable targets; and how LLMs can support misinformation detection and mitigation, and robustness-oriented countermeasures. Through this analysis, we reveal a growing mismatch between increasingly sophisticated LLM-enabled attacks and existing defenses. Future research should move toward ecosystem-level robustness by addressing three key gaps: more realistic risk evaluation under budgeted, multilingual, multimodal, competing, and long-horizon attack settings; stronger system robustness for evidence provenance, retrieval integrity, social manipulation, and secure agentic workflows; and more deployable defenses with risk-adaptive resource allocation, auditability, and post-deployment monitoring. We hope this work provides a unified foundation for understanding, evaluating, and mitigating misinformation risks in the era of LLMs.


\end{document}